\documentclass[article]{IEEEtran}
\usepackage{amsmath,amssymb,latexsym,cite}
\usepackage{epsf, pstricks,pgf, xcolor,pst-func}

\def\argmax{\operatornamewithlimits{arg\,max}}
\def\argmin{\operatornamewithlimits{arg\,min}}

\def\modl{\hspace{-0.1in}\mod\Lambda}

\def\modq{\hspace{-0.1in}\mod q}

\def\modqz{\hspace{-0.1in}\mod q\Zbb^n}

\def\Re{\mathsf{Re}}
\def\Im{\mathsf{Im}}

\def\wb{\mathbf{w}}

\def\xb{\mathbf{x}} 

\def\yb{\mathbf{y}}
\def\zb{\mathbf{z}}

\def\xr{\mathbf{x}_{\text{R}}}
\def\yr{\mathbf{y}_{\text{R}}}
\def\zr{\mathbf{z}_{\text{R}}}
\def\Xr{X_{\text{R}}}
\def\Yr{Y_{\text{R}}}
\def\Zr{Z_{\text{R}}}

\def\Fbb{\mathbb{F}}
\def\Zbb{\mathbb{Z}}

\newcommand{\snr}{\text{$\mathsf{SNR}$}}

\newtheorem{remark}{Remark}

\begin{document}

\title{Reliable Physical Layer Network Coding}
\author{Bobak Nazer, \IEEEmembership{Member, IEEE} and Michael Gastpar, \IEEEmembership{Member, IEEE}
\thanks{This work was supported by the National Science Foundation under grants CCR 0347298, CNS 0627024, and CCF 0830428 as well a Graduate Research Fellowship. } % this should stop a space
\thanks{B. Nazer is with the Department of Electrical and Computer Engineering, Boston University, Boston, MA 02215, USA (e-mail: bobak@bu.edu).} \thanks{M. Gastpar is with the Department of Electrical Engineering and Computer Sciences,
University of California, Berkeley, CA 94702, USA and with the Department of EEMCS,
Delft University of Technology, The Netherlands (e-mail: gastpar@eecs.berkeley.edu).}}

\markboth{To Appear in the Proceedings of the IEEE}{~}

\maketitle

\begin{abstract}
When two or more users in a wireless network transmit simultaneously, their electromagnetic signals are linearly superimposed on the channel. As a result, a receiver that is interested in one of these signals sees the others as unwanted \textit{interference}. This property of the wireless medium is typically viewed as a hindrance to reliable communication over a network. However, using a recently developed coding strategy, interference can in fact be harnessed for \textit{network coding}. In a wired network, (linear) network coding refers to each intermediate node taking its received packets, computing a linear combination over a finite field, and forwarding the outcome towards the destinations. Then, given an appropriate set of linear combinations, a destination can solve for its desired packets. For certain topologies, this strategy can attain significantly higher throughputs over routing-based strategies. Reliable physical layer network coding takes this idea one step further: using judiciously chosen linear error-correcting codes, intermediate nodes in a wireless network can directly recover linear combinations of the packets from the observed noisy superpositions of transmitted signals. Starting with some simple examples, this survey explores the core ideas behind this new technique and the possibilities it offers for communication over interference-limited wireless networks.
\end{abstract}
 
 
\section{Introduction}

In recent years, the number of wireless devices has skyrocketed and, to handle the demands of ever richer multimedia applications, these devices have required higher and higher data rates. These trends, coupled with the scarcity of spectrum, imply that interference between devices will be one of the dominant bottlenecks in wireless networking for many years to come. In some cases, this interference is purely an obstacle to reliable communication. However, in many scenarios, it is actually possible to harness interference to enable more efficient communication over a network. In this survey, we examine a set of novel strategies geared at exploiting wireless interference for reliable network coding.

Nodes in a network occupy one or more of the following roles: sources transmit information packets into the network, destinations are interested in recovering a set of information packets, and relays help move information between sources and destinations. In a wired network, he classical approach is to have relays forward a subset of their observed packets towards the intended destinations. For a wired network with a single source, multiple relays, and a single destination, this routing strategy is optimal \cite{ff56,efs56}. More  generally, routing cannot attain the maximum throughput and relays may need to send out functions of the packets they observe, rather than just repeating them. This \textit{network coding} strategy was originally developed by Ahlswede \textit{et al.} for optimal multicasting over wired networks \cite{acly00} and has turned out to be quite useful for a wide array of networking scenarios. Much of this research has focused on {\em linear} network coding strategies, where the functions are assumed to be linear combinations of the packets, taken over an appropriate finite field \cite{lyc03,km03}. 

In a wireless setting, transmitting a packet from one node to another naturally causes interference to all nearby nodes. If multiple nodes transmit concurrently, their waveforms are linearly superimposed which makes it harder for a receiver to recover its desired packets. Yet, for network coding, relays do not need to recover the contents of individual packets, only an appropriate functions thereof. In this overview, we will look at physical layer modulation and coding techniques that can harness the linear nature of wireless interference for linear network coding. If the relays can transmit in a fully analog fashion, one possible approach is to have them repeat their observed noisy linear combinations directly. While this has favorable properties when the signal-to-noise ratio is high enough, it is clear that the ensuing end-to-end noise accumulation is highly undesirable. A more interesting question is thus whether the noise can be removed at each stage by appropriate error-correcting codes. This is what we will refer to as {\em reliable physical layer network coding} in this paper.

Interestingly, an initial information-theoretic analysis might suggest that such coding is not feasible and instead, relays must first decode the individual packets. In other words, attempting to directly decode only a linear combination of the messages will implicitly also reveal the individual messages, and thus, not be any
more efficient than the standard approach. Fortunately, this initial attempt is too pessimistic. The key insight is that the modulation and coding strategies should share a common \textit{algebraic structure} across transmitters. More precisely, if the transmitted waveforms are points of a lattice, then every integer combination of these waveforms is itself a point of the same lattice. Therefore, receivers can efficiently decode these linear combinations with the same framework used to decode individual packets. How efficiently, depends on how closely the coefficients of the desired linear combination match the observed channel strengths and phases. 

Reliable physical layer network coding thus involves two complementary questions: {\em (i),} how to enable encoders and decoders to exploit interfering signals for efficient function computation, and {\em (ii),} at the network level, which functions to select in order to enable efficient overall information transfer. As we will demonstrate, existing modulation techniques and linear error-correcting codes can serve as building blocks for these new encoders and decoders. Information can be encoded digitally into packets at the transmitter side and decoded directly into a linear combination of packets at the receiver side. We will illustrate the basic ideas behind this new approach starting with very simple examples and gradually incorporating many of the aspects of wireless channels. In keeping with the survey nature of this paper, we will point to relevant papers in the literature along the way.

\section{Network Coding Preliminaries}

Consider a network of several nodes, some of which are linked together by wired connections. If one node wishes to send a message to another node in the network, then it is optimal to simply route the information towards its destination: intermediate nodes should simply retransmit their received packets. This strategy can achieve the \textit{unicast capacity} of a network which is given by the max-flow min-cut theorem, as shown independently by Ford and Fulkerson \cite{ff56} and Elias, Feinstein, and Shannon \cite{efs56}. Now, suppose that more than one destination wants the transmitted message. The seminal paper of Ahlswede {\em et al.} demonstrated that routing is insufficient for this problem and network coding is, in general, required to achieve the \textit{multicast capacity} \cite{acly00}. The key principle underlying network coding is that intermediate nodes should send out functions of their received packets, instead of the packets themselves. Subsequent work by Li, Cai, and Yeung \cite{lyc03} and K\"otter and M\'edard \cite{km03} made the important observation that, for multicasting, intermediate nodes can simply send out a linear combination of their received packets. There is now a wealth of literature on the myriad applications of network coding to sending information over networks and beyond. A comprehensive literature survey is beyond the scope of this paper and we refer the interested reader to the other papers in this issue as well as to several books on the subject \cite{yeungcailizhang,fragoulisoljanin1,fragoulisoljanin2,yeung,holun}.

For the purpose of our exposition and to make the ideas behind physical layer network coding apparent, we need to develop network coding slightly more formally here. In particular, we will consider operations on a finite field, i.e., a set of $q$ elements that we will denote without loss of generality by $\{0,1,2,\ldots, q-1\}.$ For ease of exposition, we will assume that $q$ is a prime number so that addition and multiplication over the finite field can be written as modulo addition and multiplication over the reals. For any two integers $a$ and $b$ in this set, we will denote addition and multiplication modulo-$q$ as
\begin{align}
a \oplus b &= [ a + b ] \modq \\
a \otimes b & = [a b ] \modq \ . 
\end{align}

We will work with the algebraic network coding framework introduced in \cite{km03}. The transmitting terminal has a message which can be represented as a string of bits. This message can be broken up into several packets each of which can be written as a length-$k$ vector of elements from the finite field which we will denote by $\mathbf{w}_\ell \in \Fbb_q^k$. Say an intermediate node (or relay) in a network has received some of these packets $\mathbf{w}_1, \mathbf{w}_2, \ldots, \mathbf{w}_L$. The node's role in a network coding solution is to send a linear combination $\mathbf{u}$ of these packets towards the destination:
\begin{align}
\mathbf{u} = a_1 \mathbf{w}_1 \oplus a_2 \mathbf{w}_2 \oplus \cdots \oplus a_L\mathbf{w}_L  \label{e:netcod}
 \end{align} where $a_1,a_2, \ldots,a_L$ are coefficients over the finite field. The goal is for each destination to collect enough linear combinations to infer the original packets. Assume that a destination has successfully received linear combinations $\mathbf{u}_1, \mathbf{u}_2,\ldots, \mathbf{u}_M$ where
\begin{align}
\mathbf{u}_m = a_{m1} \mathbf{w}_1 \oplus a_{m2} \mathbf{w}_2 \oplus \cdots \oplus a_{mL}\mathbf{w}_{L}. 
\end{align} Then, it can solve for the original packets if the matrix of coefficients
\begin{align}
\mathbf{A} = 
\left[
\begin{array}{cccc}
  a_{11}& a_{12}  & \cdots & a_{1L}  \\
  a_{21}& a_{22}  & \cdots & a_{2L}  \\
  \vdots & \vdots  & \ddots & \vdots \\
  a_{M1} &a_{M2} & \cdots &a_{ML}  
\end{array}
\right]
\end{align}has rank $L$. Since we would like all destinations to recover the message, we must choose the network coding coefficients at each relay so that the matrix of coefficients at each receiver is full rank. Jaggi \textit{et al.} developed an efficient algorithm that can find a feasible set of coefficients in polynomial time so long as the field size $q$ is larger than the number of receivers \cite{jsceej05}. Another powerful approach advocated by K\"otter and M\'edard as well as Ho \textit{et al.} is to generate the coefficients randomly at each relay \cite{km03, hkmesk06}. It can be shown that the probability that this yields a valid solution increases with the field size.

It is also instructive to note that within this framework, the routing solution corresponds to having the intermediate node retransmit one of its received packets,
 \begin{align}
 \mathbf{u} = \mathbf{w}_\ell ~~~ \mbox{for some } \ell \in \{1,2,\ldots,L\} \ ,
 \end{align} during each time slot.
 
\subsection{Two-Way Relay Channel}\label{s:twoway}
 
\begin{figure}[h]
\centering
\includegraphics[width=3.25in]{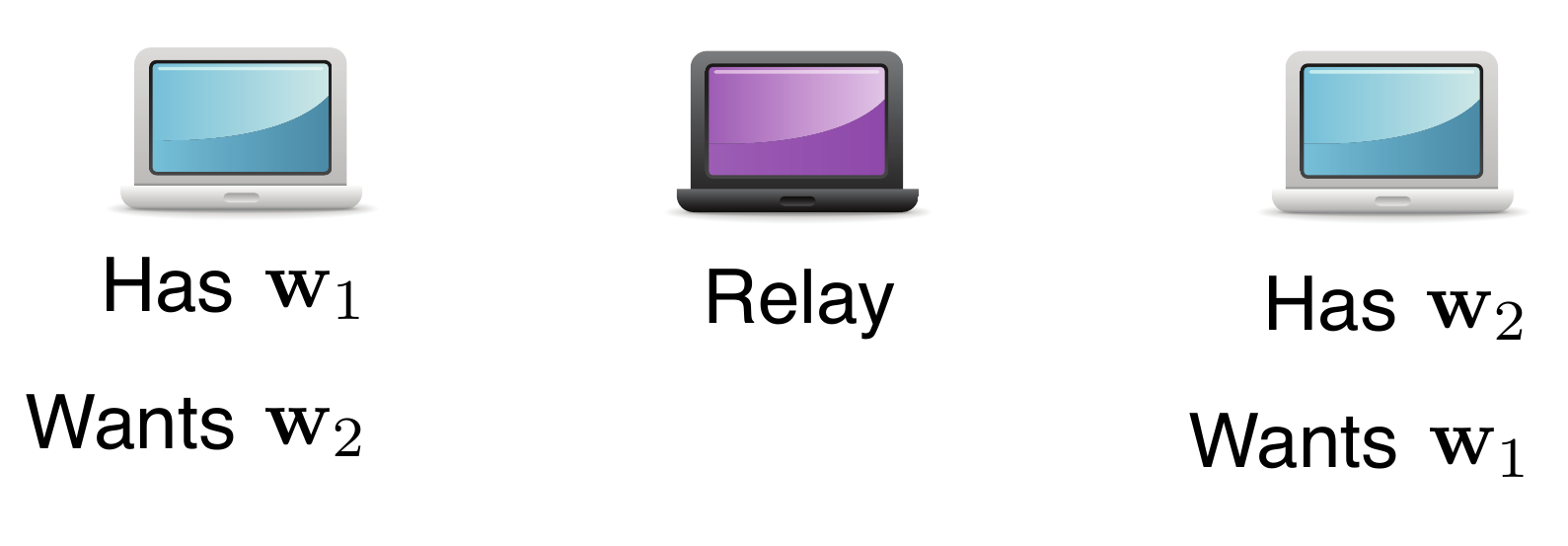}
\caption{Two-way relay channel.}\label{f:twowayrelay}
\end{figure}

We now introduce a simple network, \textit{the two-way relay channel}, that will serve as a guiding example and benchmark for all of the strategies in the sequel. To the best of our knowledge, this example first appeared in a paper by Wu, Chou, and Kung in 2004 \cite{wck04}. As shown in Figure \ref{f:twowayrelay}, there are two users that wish to exchange messages with each other. However, in this model, we assume that each user cannot hear the other user's transmission. Instead, the users must communicate with the help of a relay node ``in the middle'' that can receive from and transmit to both users. 

\begin{figure}[h]
\centering
\includegraphics[width=3.25in]{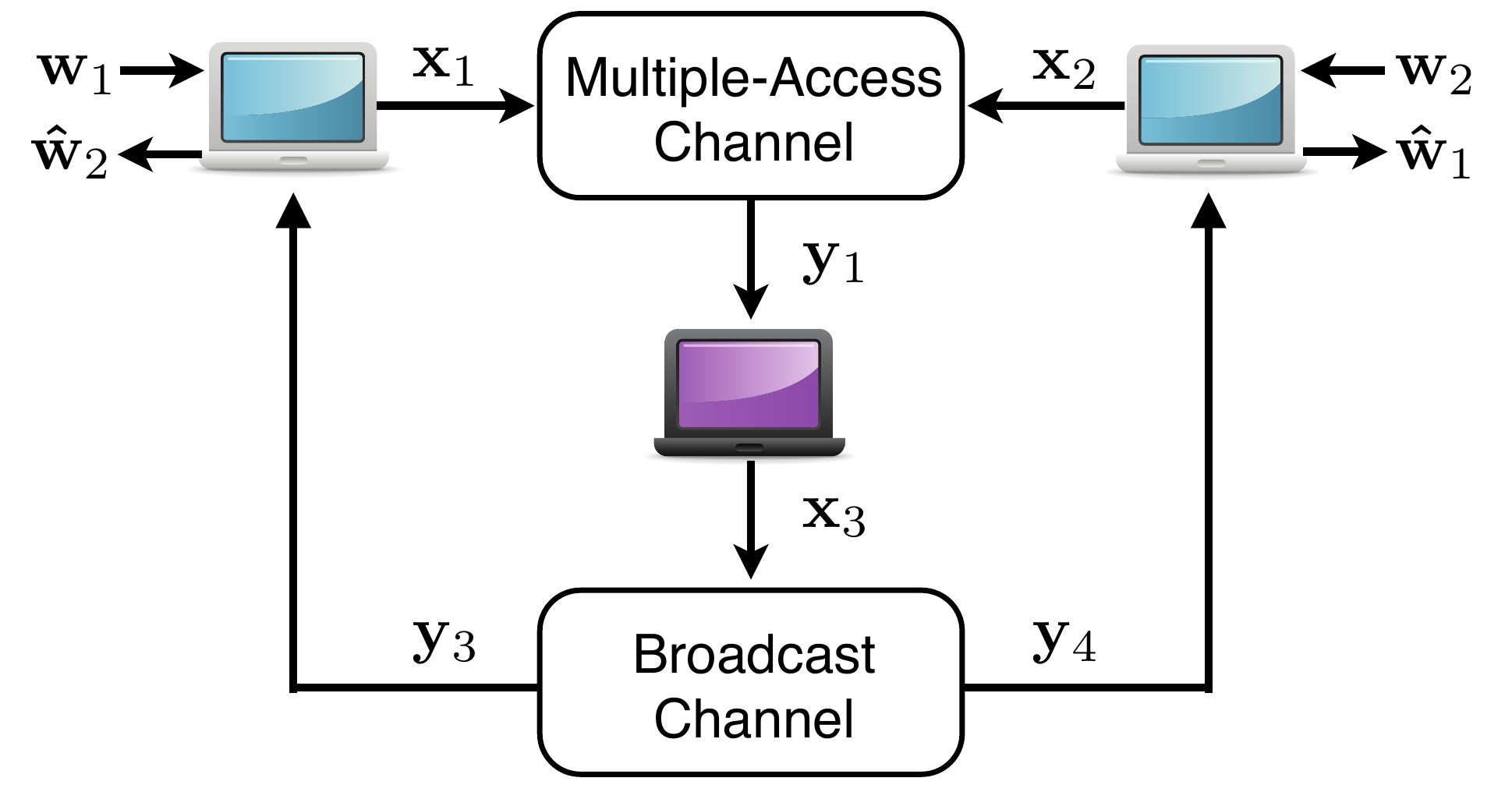}
\caption{Two-way relay channel model.}\label{f:twowaychanmodel}
\end{figure}

The users share the same frequency band so, if both users transmit simultaneously, the relay will observe a superposition of the two signals corrupted by noise. This interference effect can be modeled by a \textit{multiple-access channel} with inputs $\mathbf{x}_1$ and $\mathbf{x}_2$ and output $\yr$. Each user $m=1,2$ generates its channel input $\mathbf{x}_m$ from its message $\mathbf{w}_m$ and the channel output $\mathbf{y}_1$ is observed by the relay (see Figure \ref{f:twowaychanmodel}). Conversely, any transmission by the relay will be heard by both users. This broadcast effect can be modeled by a \textit{broadcast channel} with input $\xr$ and outputs $\mathbf{y}_1$ and $\mathbf{y}_2$. 

These two channel models have been studied in depth over the past few decades. From an information-theoretic perspective, the capacity region for sending messages over a multiple-access channel has been completely characterized \cite{liaophd, ahlswede71}. The capacity region of the broadcast channel is also known if the channel is ``stochastically degraded'' \cite{cover72,bergmans74,gallager74}. This condition holds for the wireless channel models considered in this paper. We will not delve into the subtleties of these capacity results and refer the interested reader to \cite{coverthomas,kramer,elgamalkim} for more details. Roughly speaking, these results tell us that for wireless multiple-access and broadcast channels at symmetric operating points, each user can attain a rate inversely proportional to the number of active users. For the two-way relay channel, this means that sending two messages to the relay takes approximately twice as much time as sending one message. Similarly, sending two different messages from the relay to two destinations takes twice as much time as sending one message to one destination. 

\begin{figure}[h]
\centering
\includegraphics[width=3.25in]{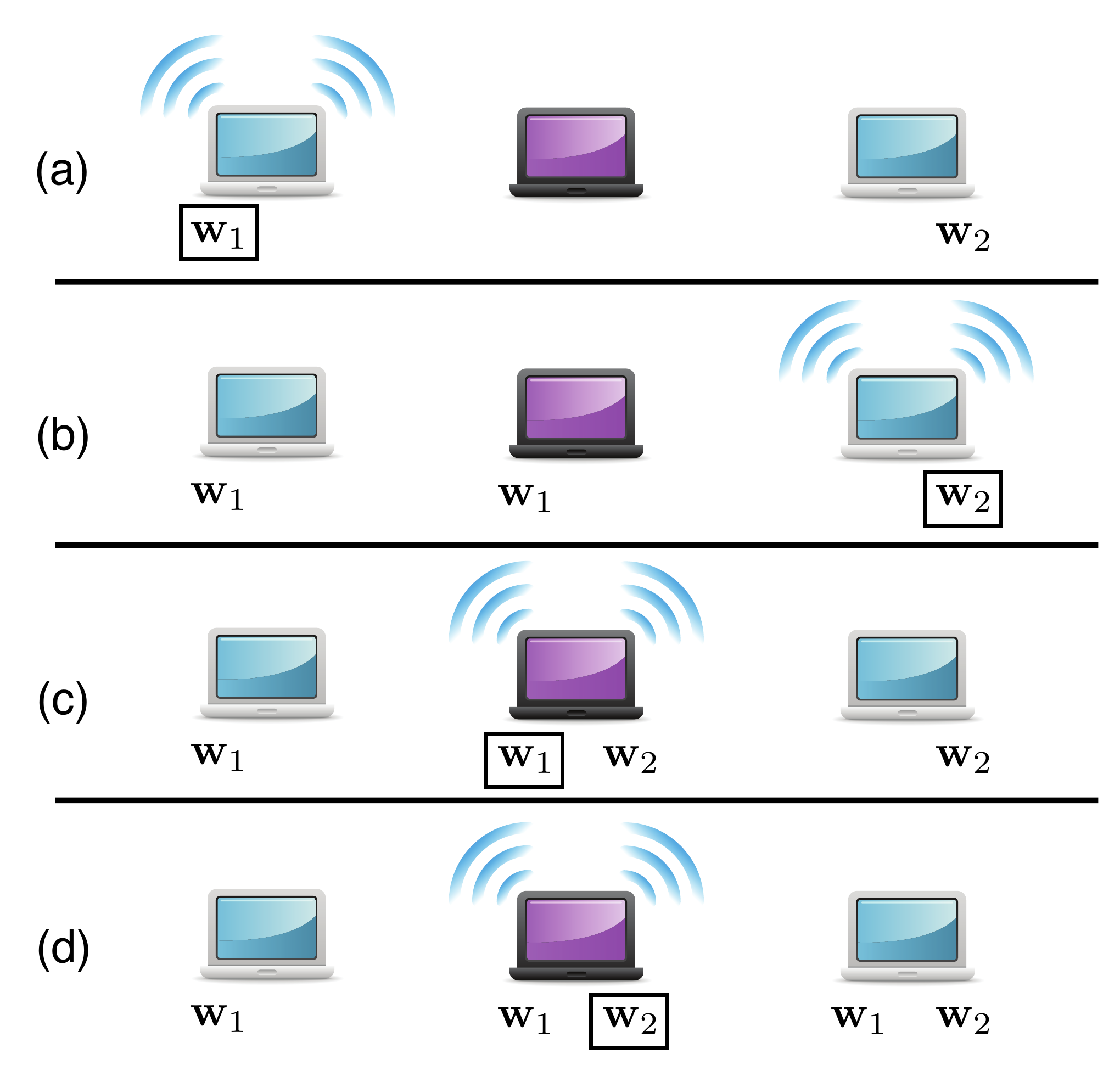}
\caption{A routing strategy for the two-way relay channel that requires $4$ time slots. (a) During the first time slot, user $1$ sends its message $\mathbf{w}_1$ to the relay. (b) During the second time slot, user $2$ sends its message $\mathbf{w}_2$ to the relay. (c) During the third time slot, the relay sends the message $\mathbf{w}_1$ to user $2$. (d) During the fourth time slot, the relay sends the message $\mathbf{w}_2$ to user $1$. }\label{f:twowayrouting}
\end{figure}

We make the natural assumption that each terminal must operate in half-duplex mode (i.e. it can either send or receive during a single time slot but not both). From this, we get that combining standard physical layer coding ideas with routing allows both users to exchange messages over the two-way relay channel in $4$ time slots. We illustrate how this can be done in Figure \ref{f:twowayrouting}. Each user takes a time slot to send its message to the relay and the relay takes two time slots to send these messages to their destinations.

\begin{figure}[h]
\centering
\includegraphics[width=3.25in]{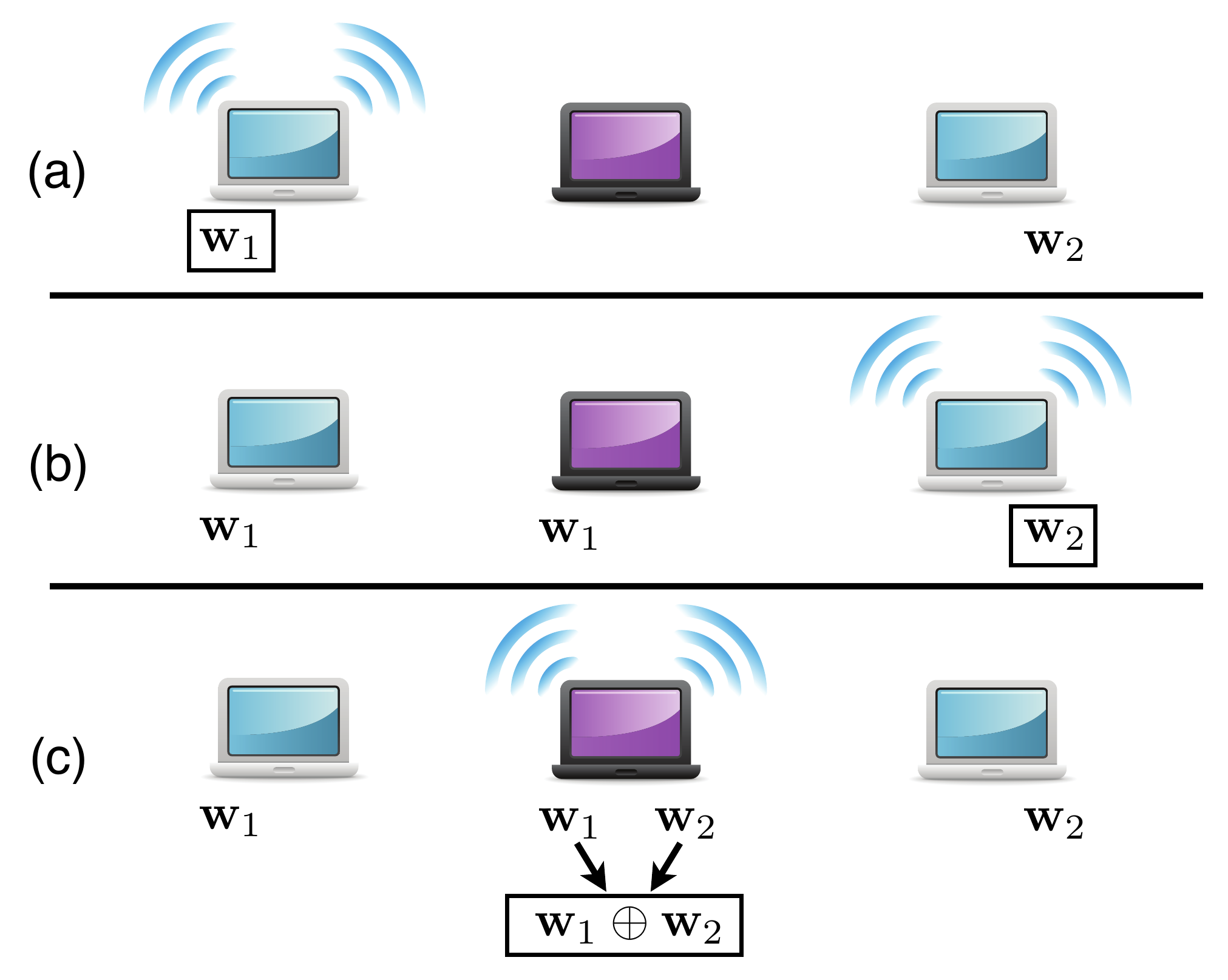}
\caption{A network coding strategy for the two-way relay channel that requires $3$ time slots. (a) During the first time slot, user $1$ sends its message $\mathbf{w}_1$ to the relay. (b) During the second time slot, user $2$ sends its message $\mathbf{w}_2$ to the relay. (c) During the third time slot, the relay sends the sum of the messages $\mathbf{w}_1 \oplus \mathbf{w}_2$ to both users. }\label{f:twowaynetcod}
\end{figure}

This performance can be significantly improved through the use of network coding \cite{wck04}. Once the relay has collected $\mathbf{w}_1$ and $\mathbf{w}_2$ (after two time slots) it can easily compute the modulo-$2$ sum of the messages $\mathbf{w}_1 \oplus \mathbf{w}_2$. In the next time slot, it can broadcast the sum to both users. Each user can then infer its desired message from the sum and its original message. This network coding strategy thus allows the users to exchange messages in only $3$ time slots as shown in Figure \ref{f:twowaynetcod}. This gain is not in conflict with the capacity region of the broadcast channel as the relay only needs to send out one common message rather than two distinct messages. 

\begin{remark} Broadcasting  a common message is limited by the weakest channel from the relay to a single user. We also note that in general, it is not optimal to send a common message comprised of the {\em sum} of the bits \cite{tuncel06,xie07,ks07,osbb08}, but, for the channel models we will consider, it will suffice.
\end{remark}

The two-way relay channel is just one of the many scenarios where the broadcast property of the wireless medium can be exploited via network coding. This behavior has been thoroughly investigated theoretically \cite{wck04,dehkklmr05,lrmkkhaz06,dgphe06,rk06,sv06,ljs06,wu07,kmt08,osbb08,zl09,lgt09} and demonstrated in practice \cite{krhkmc08}.

\begin{figure}[h]
\centering
\includegraphics[width=3.25in]{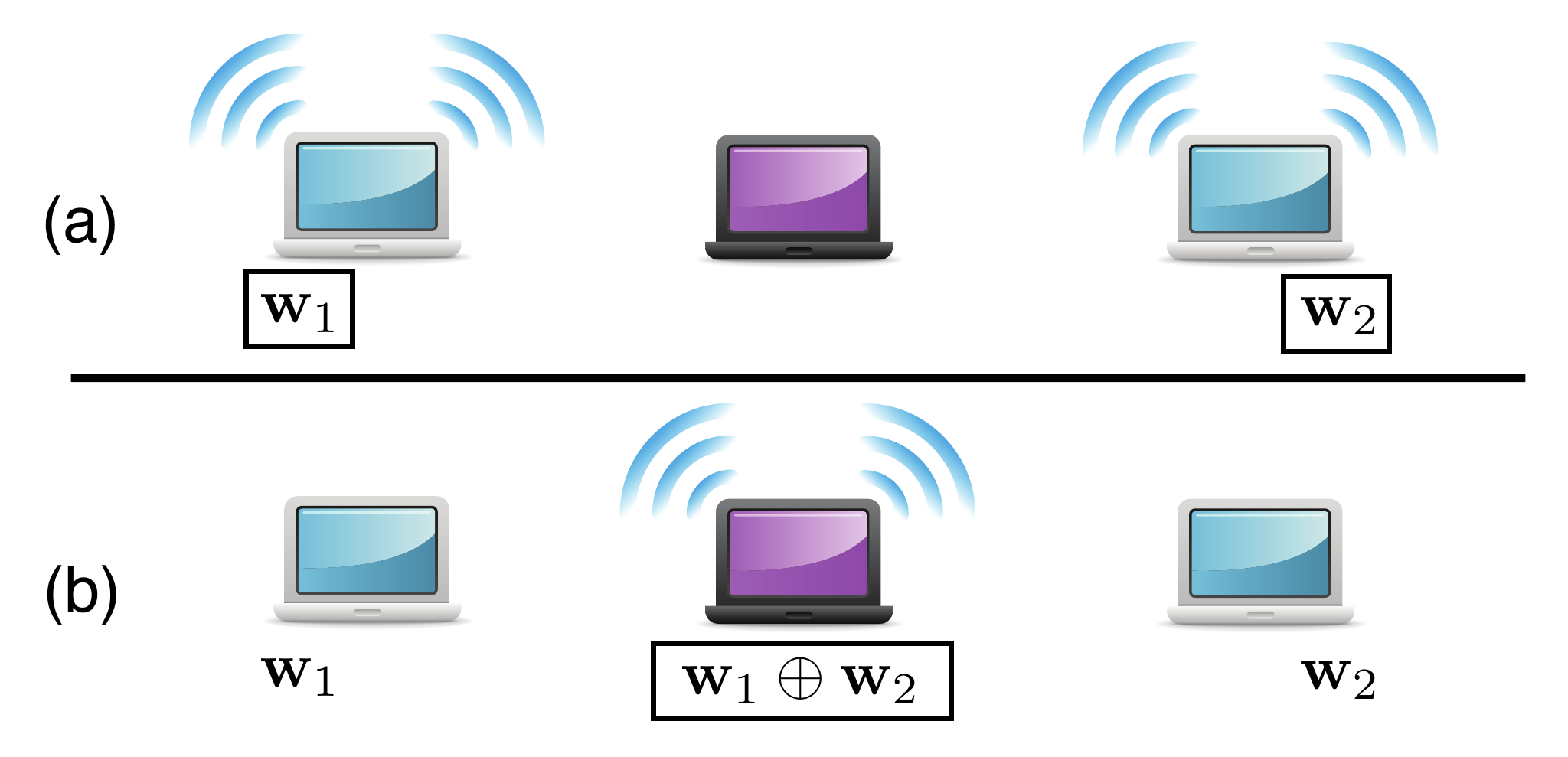}
\caption{A physical layer network coding strategy for the two-way relay channel that requires $2$ time slots. (a) During the first time slot, the users send the sum of their messages $\mathbf{w}_1 \oplus \mathbf{w}_2$ to the relay using a nested lattice scheme. (b) During the second time slot, the relay sends the sum of the messages back to both users. }\label{f:twowaycomp}
\end{figure}

Now that we know it is more efficient for the relay to send the sum of the messages during the broadcast phase, it is natural to ask whether savings are also possible during the multiple-access phase. Since the relay only needs the sum, we could do even better by conveying the sum to the relay without identifying the individual messages. Simultaneously transmitted signals are added up on the wireless channel and, as we will show, this property can be exploited to send the sum (or another linear function) to the relay in a single time slot. Figure \ref{f:twowaycomp} illustrates how this scheme allows users to exchange messages in just $2$ time slots. The remainder of this paper is devoted to showing how this is possible and characterizing the exact gains. We will start with some rudimentary examples and gradually build up a toolkit for general networks.

\begin{remark} Note we did not take into consideration the distributed scheduling problems that arise in wireless networks. For instance, in the backoff protocol of the IEEE 802.11 standard, users listen to see if the channel is free before they transmit. If the channel is in use, they remain silent for a random interval before listening again. We are primarily interested in the gains due to novel signaling schemes and will assume ideal scheduling. The interaction between these new schemes and practical scheduling algorithms is beyond the scope of this paper. 
\end{remark}

\section{A Finite Field Physical Layer} \label{s:finitefield}

We start our discussion by considering a hypothetical physical layer that is particularly well-suited
to the standard linear network coding, yet serves to illustrate some of the key properties and effects
that can be exploited over more realistic physical layers, such as the wireless case discussed
in the second part of this paper. This finite field model will help build intuition for the more intricate strategies used later on.

\subsection{Noiseless Interference}

Specifically, let us first consider a channel model that is completely noise-free,
and where the transmitted signals interfere in a modulo-additive way.
That is, each transmitted symbol takes values in $\{0,1,2,\ldots,q-1\},$
where we assume that $q$ is a prime number. Let $x_\ell[t]$ denote the symbol transmitted by the $\ell^{\text{th}}$ user in time slot $t$. Separately for each time slot, the physical layer provides, as its channel output, the modulo-$q$ sum of all the input signals from all $L$ transmitters during that time slot:
\begin{align}
y[t] = x_1[t] \oplus x_2[t] \oplus \cdots \oplus x_L[t] \ .
\end{align}
We find it convenient to consider blocks of $n$ time slots jointly,
which we represent in vector notation:
\begin{align}
\mathbf{x}_\ell &= \big[ x_\ell[1]~x_\ell[2]~\cdots~x_\ell[n] \big]^T\\
\mathbf{y} &= \big[ y[1]~y[2]~\cdots~y[n] \big]^T
\end{align} where $^T$ is the transpose operator.
Thus, the channel output can be expressed as
\begin{align}
\mathbf{y} = \mathbf{x}_1 \oplus \mathbf{x}_2 \oplus \cdots \oplus \mathbf{x}_L \ .
\end{align}
It is quite simple to exploit this channel for linear network coding. Each transmitter should just pre-multiply its packet by an appropriately chosen coefficient $a_\ell$ out of the set $\{0,1,2,\ldots,q-1\},$ where multiplication is again modulo-$q,$ and transmit the resulting signal on the physical layer,
\begin{align}
\mathbf{x}_\ell = a_\ell \mathbf{w}_\ell \ .
\end{align}
The channel output is then exactly equal to our desired linear combination
\begin{align}
\mathbf{y}_\ell &= a_1 \mathbf{w}_1 \oplus a_2 \mathbf{w}_2 \oplus \cdots \oplus a_L \mathbf{w}_L \ . 
\end{align}
In one shot, the receiver learned exactly what it needed to make the network code work and not one bit more. This should be contrasted to the standard approach in which the transmitters would take turns, each sending
its entire packet to the receiver, who would then compute the desired linear combination (and forward it).
Clearly, the latter would take $L$ times longer to complete. Therefore, for this very particular ``physical'' layer, the described (simple and obvious) scheme attains a speedup of a factor of $L,$ which can also be shown to be the optimal attainable performance (see Section \ref{s:optimality}). It will be convenient to capture the performance of the computation scheme by what we will refer to as the {\em computation rate,} namely, the number of bits of the linear function successfully recovered per channel use. For the present example, this evaluates to
\begin{align}
R_{\text{COMP}} =  \log_2{q}
\end{align} whereas sending the data separately can only attain
\begin{align}
R_{\text{COMP}} = \frac{1}{L} \log_2{q} \ .
\end{align}

\begin{figure}[h]
\centering
\includegraphics[width=3.5in]{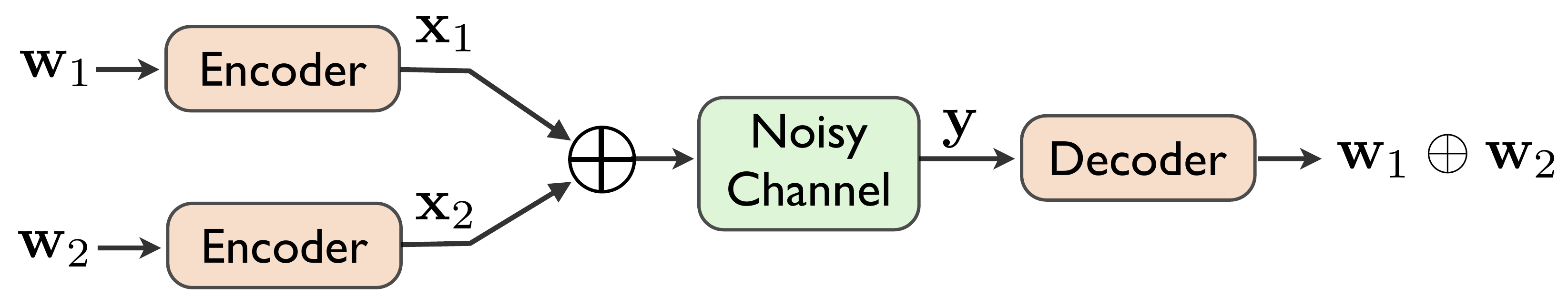}
\caption{A noisy modulo-adder channel.}\label{f:modadder}
\end{figure}

\subsection{Interference with Erasures}

\begin{figure}[h]
\centering
\includegraphics[width=3.5in]{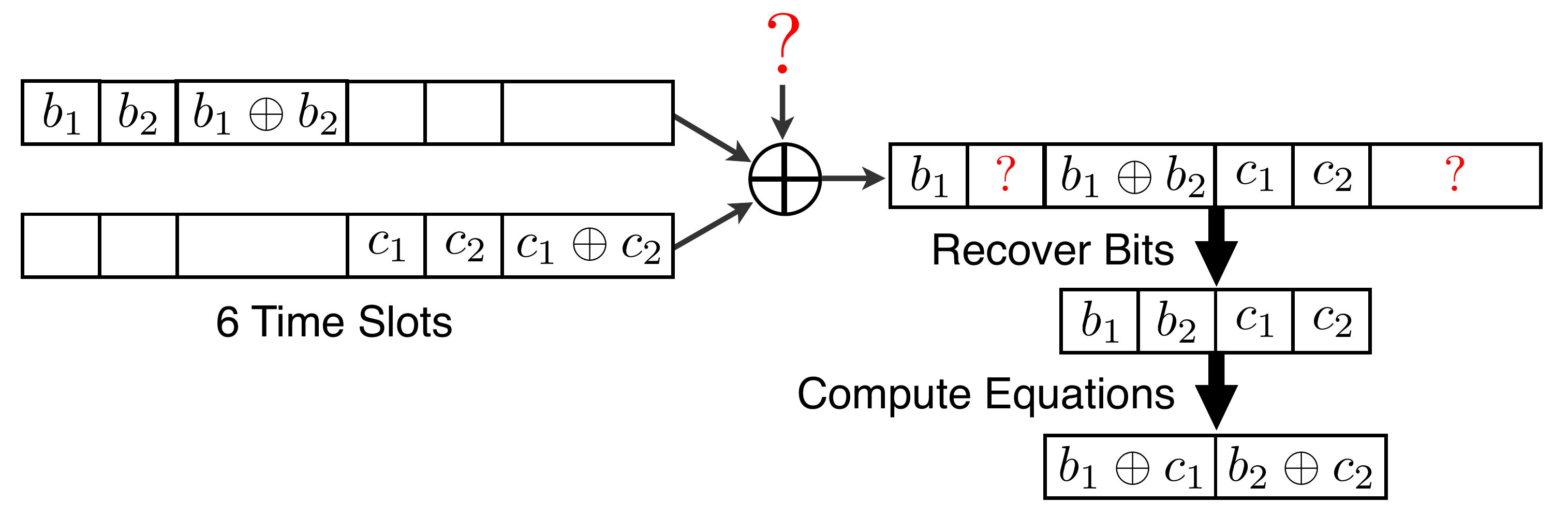}
\caption{Reliable computation over a modulo-adder with erasures. Users take turns sending their data. Afterwards, the receiver computes the desired sum. }\label{f:erasuresep}
\end{figure}

\begin{figure}[h]
\centering
\includegraphics[width=3.5in]{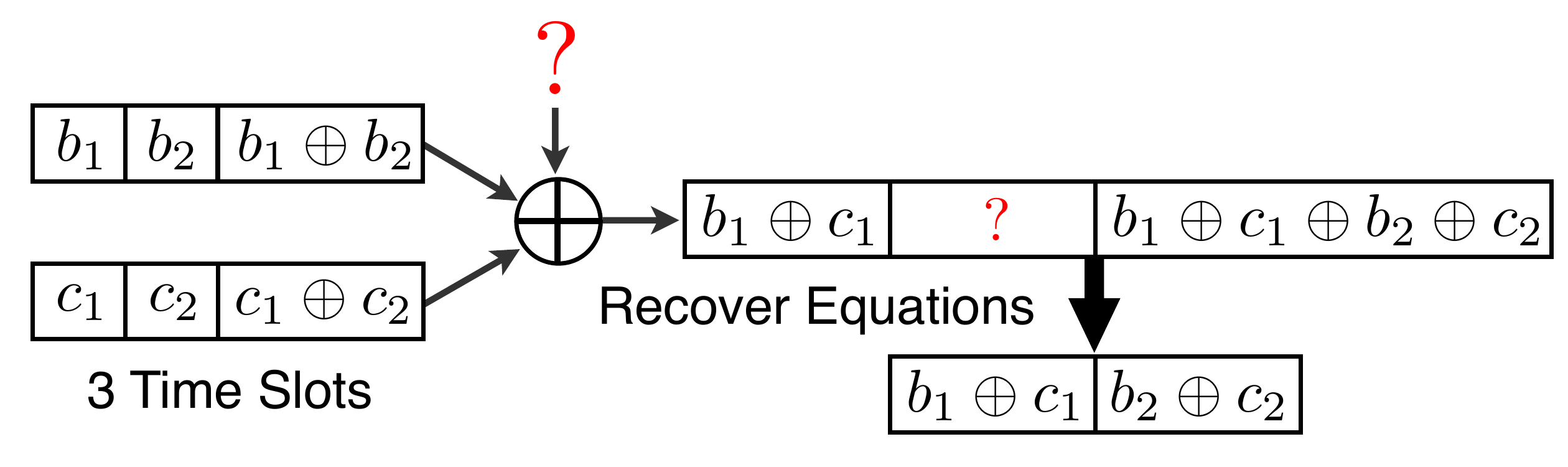}
\caption{Reliable computation over a modulo-adder with erasures. Users send their data at the same time and the receiver directly infers the desired sum.}\label{f:erasurecomp}
\end{figure}

Let us now bring this model one step closer to physical reality. In particular, we will add {\em noise}
into the picture and show that in some interesting cases, exactly the same speedup is possible.
We begin with an erasure channel. Say that out of every block of three symbols put into the channel, one is chosen at random and erased. If we just transmit uncoded, some of the symbols will be lost and we will fail to achieve our goal of error-free network coding. We can overcome this problem with an error-correcting code that adds a parity-check to every two symbols. Let $b_1$ and $b_2$ be symbols taking values in $\{0,1,2,\ldots,q-1\}$. The code maps these two information symbols into three symbols
\begin{align}
[b_1~~b_2~~b_1 \oplus b_2]
\end{align} which can then be transmitted over the channel. If one of these three symbols is missing, it can be recovered from the other two so we can reliably transmit symbols over the channel. Now, consider a two-user channel like that shown in Figure \ref{f:modadder}. The output is the mod-$q$ sum of the two users' transmissions, except that one out of three symbols is randomly erased. In Figure \ref{f:erasuresep}, we illustrate a standard approach to error control over a noisy multiple-access channel: each user encodes its own symbols and is allocated its own time slots for transmission. Our ultimate goal is to reconstruct the linear combinations $b_1 \oplus c_1$ and $b_2 \oplus c_2$ at the decoder. Transmitting all the symbols reliably to the receiver requires a total of $6$ channel uses as we need one parity symbol per user to recover from the erasure.  However, as shown in Figure \ref{f:erasurecomp}, if both users transmit their coded symbols simultaneously, this can be accomplished in $3$ channel uses. The parity checks $b_1 \oplus b_2$ and $c_1 \oplus c_2$ will be added up by the channel. Therefore, the receiver will observe $b_1 \oplus c_1 \oplus b_2 \oplus c_2$, which serves as a parity check on the desired linear combinations. In other words, the channel combines the original parity checks in exactly the right way. In fact, if all the information was available at a single transmitter, we would pre-compute the linear combinations and use the same parity check to protect them. In summary, the key observation is a speedup (or, equivalently, capacity gain) proportional to the number of transmitting terminals (two, in the example just discussed). More explicitly, the computation rate for this scheme is
\begin{align}
R_{\text{COMP}} =  \frac{2}{3} \log_2{q}
\end{align} and the computation rate resulting from sending all of the data is 
\begin{align}
R_{\text{COMP}} = \frac{1}{3} \log_2{q} \ .
\end{align}

\subsection{Interference with Modulo-Additive Noise}

The key idea in the example above is that not only does the channel naturally compute the linear combinations of the information symbols, it also computes the parity checks for them. As it turns out, this idea can take us quite far for noisy modulo-adder channels. 

\subsubsection{Algebraic Coding Perspective}
We find it instructive to start the discussion by considering classical algebraic error-correction codes.
To this end, let $\mathbf{G}$ be an $n \times k$ generator matrix for a linear code that can correct up to $d$ errors. Both users in Figure \ref{f:modadder} encode their respective messages using this generator matrix,
meaning that they will choose channel inputs $\mathbf{x}_1 =  \mathbf{G}\mathbf{w}_1$ and $\mathbf{x}_2 =\mathbf{G}\mathbf{w}_2$, respectively.
Now, the the channel output can be expressed as
\begin{align}
\mathbf{y} &= \mathbf{x}_1 \oplus \mathbf{x}_2 \oplus \mathbf{z} \\
&= \mathbf{G} \mathbf{w}_1 \oplus \mathbf{G} \mathbf{w}_2 \oplus \mathbf{z} \\
&= \mathbf{G}(\mathbf{w}_1 \oplus \mathbf{w}_2) \oplus \mathbf{z}
\end{align} 
where the last step follows from the distributive property of matrix multiplication. The key observation is that the generator matrix $\mathbf{G}$ directly protects the modulo-sum of the two messages. Therefore, as long as there are no more than $d$ errors, i.e., as long as the Hamming weight\footnote{The Hamming weight of a vector is the number of symbols that are not equal to zero.} of the error vector $\mathbf{z}$ is no more than $d,$ the modulo-sum of the messages can be perfectly recovered. It is important to note that this again represents a speedup of a factor proportional to the number of users (two, in this example) over the standard approach of sending the full messages to the decoder.

\subsubsection{Information-Theoretic Perspective} \label{s:optimality}

Next, we discuss how this idea naturally extends beyond the fixed error model of classical error-correction codes to models with random errors. To begin, it is insightful to consider the standard connection between
such codes and classical information theory. The standard information-theoretic approach for showing 
the achievability of a certain rate of transmission is via the so-called random coding argument, a version of the probabilistic method. First, codewords are drawn element-by-element independently from a fixed probability
distribution. Then, it is shown that the decoding error probability at the receiver is very small as long as the number of codewords is small enough. This involves a union bound over all codewords. Finally, it can be argued that since the probability of error is small on average over the random codebook, there must be at least one good fixed codebook (see \cite[Theorem 7.7.1]{coverthomas} for more details). Clearly, the resulting randomly chosen codebook used in this argument has no algebraic structure whatsoever (with probability one).

However, there is an alternative argument that also permits to establish achievable rates of communication and that does involve algebraic structure. Here, one starts by randomly drawing each element a generator matrix $\mathbf{G}$  independently from the uniform distribution over $\{0,1,2,\ldots,q-1\}.$ We note that by this construction, the elements of all codewords will be uniformly distributed over the entire $q$-ary alphabet.
It can be shown with a little more work that the resulting code has pairwise independent codewords (see \cite[Section 6.2]{gallager}) so one can still take a union bound as above.
The main trick is to
let each transmitter employ this same code:
\begin{align}
\mathbf{x}_\ell = a_\ell \mathbf{G} \mathbf{w}_\ell \ . \label{e:samelinear}
\end{align}
The channel output then looks as if the desired function $\mathbf{u}$ was directly encoded with $\mathbf{G}$:
\begin{align}
\mathbf{y} &=  \mathbf{G}\left(a_1 \mathbf{w}_1 \oplus \cdots \oplus a_L \mathbf{w}_L \right) \oplus \mathbf{z}\\ &=  \mathbf{G}\mathbf{u} \oplus \mathbf{z} \ .
\end{align}
Since the codewords are pairwise independent, standard information-theoretic arguments
can be applied. It can be shown that the probability of decoding error can be made
arbitrarily small (by increasing the coding blocklength) so long as the rate is smaller than
the mutual information between the sum\footnote{One intriguing mathematical issue that arises here is that the usual proof techniques, such as random coding, are not able to take advantage of the channel between the sum of the inputs and the output: algebraically structured codes seem to be necessary \cite{ng08ETT}.}  of the inputs $X_1 \oplus \cdots \oplus X_L$ and the output $Y,$
which can be calculated as follows:
\begin{align}
&I(X_1\oplus \cdots \oplus X_L; Y) \\&~~~~= H(Y) - H(Y| X_1 \oplus \cdots \oplus X_L) \\&~~~~= H(Y) - H(X_1 \oplus \cdots \oplus X_L \oplus Z| X_1 \oplus \cdots \oplus X_L) \nonumber \\ &~~~~= H(Y) - H(Z) \ . 
\end{align} 
For this random linear code construction, the channel inputs are uniformly distributed over all $q$ letters,
and thus, $Y$ is also uniformly distributed over all $q$ letters, meaning that
$H(Y) = \log_2q.$ Thus, any rate up to  
\begin{align}
R_{\text{COMP}} = \log_2{q} - H(Z)
\end{align} bits per channel use is achievable. For comparison, sending all of the data separately and then evaluating the function requires $L$ times more channel uses, resulting in a rate 
\begin{align}R_{\text{COMP}} = \frac{1}{L}(\log_2{q} - H(Z)) \ .
\end{align} bits per channel use.

The idea of using the \textit{same linear code} at each encoder for this setting was introduced by us in a 2005 paper \cite{ng05} (see \cite{ng07IT} for the journal version). Our inspiration came from a paper by K\"orner and Marton which uses this technique for the distributed compression of the parity of two dependent binary sources \cite{km79}. The gains in their setting come from the degree of dependence between the two binary random variables. In our case, the gains come from eliminating the need to send all the data to the receiver.\footnote{In fact, we can also take advantage of the dependencies between messages while exploiting the channel's natural computation. To simplify the presentation, we have assumed throughout that users' messages are independent. See \cite{ng07IT} for more details.}

We now briefly discuss the information-theoretic optimality of the proposed coding technique for the
special case of linear modulo-additive channel models. For such channels, the rate cannot exceed \begin{align}
R & \le \max_{p(x_1,\ldots,x_L)} I(X_1,\ldots,X_L; Y) 
\end{align} as this is the best performance attainable if all transmitters could fully cooperate. Note that the maximization is over the probability distribution of the channel inputs
$(X_1, X_2, \ldots, X_L).$ Rewriting this mutual information expression in terms of entropies yields
\begin{align}
&I(X_1,\ldots,X_L;Y)  \\
&=H(Y) - H(Y|X_1, \ldots, X_L) \\
&=H(Y) - H(X_1 \oplus \cdots \oplus X_L \oplus Z|X_1, \ldots, X_L) \\
&=H(Y) - H(Z) \\ 
&\leq \log_2{q} - H(Z) \ . 
\end{align} where the inequality is due to the fact that the entropy cannot be larger than the logarithm of the alphabet size. This means that in the special case of the modulo-additive channel,
the proposed code attains the best possible computation rate,
i.e. the computation capacity.\subsection{Beyond Finite Field Models}

This coding strategy can certainly be employed in channels that cannot be represented as a noisy finite field sum of their inputs. Although it may not always attain the capacity, it often provides a superior performance than sending all of the data to the receiver. Assume each channel input takes values on $\{0,1,2,\ldots,q-1\}$ (otherwise relabel the symbols appropriately). We use the same encoding procedure as before by having each transmitter send $\mathbf{x}_\ell = a_\ell \mathbf{Gw}_\ell$. It can be shown that the receiver can reliably decode the function $a_1 \mathbf{w}_1 \oplus \cdots \oplus a_L \mathbf{w}_L$ so long as the rate is less than the mutual information between this desired function and the channel output:
\begin{align}
R_{\text{COMP}} < I(a_1 X_1 \oplus \cdots \oplus a_L X_L; Y)
\end{align} where $X_1, \ldots, X_L$ are uniformly distributed. See \cite[Theorem 7]{nazerphd} for more details. Another possible strategy is to have each transmitter send its data uncoded simultaneously and then take turns sending parity checks while avoiding collisions. In some cases, this does better than using the same linear code at each transmitter. We studied this strategy in more depth in \cite{ng07IT,nazerphd} under the moniker of ``systematic computation coding.''

While channels that operate over the finite field may seem a bit contrived, they can be quite useful to model certain wireless scenarios. For instance, in Section \ref{s:uncodedfinite}, we will explore a scheme in which the receiver makes a hard decision on the modulo-$2$ sum of the transmitted bits. There is some probability, depending on the signal-to-noise ratio, that this estimate of the sum is in error. Thus, after the hard decision, the channel is precisely a noisy modulo-$2$ adder channel and the coding techniques developed above can be used to denoise the modulo-$2$ sum of the bits.

Finite field models can also be used to create accurate approximations of certain classes of wireless networks as shown by Avestimehr, Diggavi, and Tse \cite{adt09}. Several groups have studied network coding strategies  in the context of finite field physical layer models including \cite{emhrkkh03,ng06,bgs06,ncl09,gsgps09}.

\section{The Wireless Medium} \label{s:wireless}

The structure and interpretation of the physical layer is of key importance for the ideas presented in this paper. For example, in Section~\ref{s:finitefield}, we considered a finite field ``physical'' layer that could be easily incorporated into the network code construction, leading to significant speedup (or, equivalently, capacity gains). Much of the remainder of this paper is devoted to a physical layer that is of particular current interest, namely, the wireless medium. Therefore, we first briefly revisit its standard models and properties. We will refrain from a deep discussion of this, referring the reader instead to a host of textbooks and monographs on the topic, including~\cite{goldsmith,tseviswanath}. However, there are three key observations that are important for the techniques discussed in this paper, and we here review them in turn.

\subsubsection{Signal fading is linear}
Between the transmitter and a receiver, an electromagnetic signal undergoes a
(potentially time-varying) {\em linear} transformation.
This transformation is primarily induced by reflections
and multi-path propagation.
Assuming band-limited communication, the respective signals can be
represented uniquely by complex-valued discrete-time samples.
Then, the received signal at any point in space can be expressed as the
convolution of the transmitted signal with an impulse-response function
that characterizes the signal propagation.
Short of modeling exactly the physical surroundings, a popular
approach is to model this impulse-response function in a statistical
fashion. Assuming flat (meaning frequency non-selective) fading, as would
be appropriate for narrow-band communication, this impulse-response
function reduces to a delta function whose height characterizes
the signal propagation.
In this particularly simple and commonly studied model,
the induced signal at any point in space can be expressed as $h X[t],$
where $X[t]$ is the signal transmitted in time slot $t.$
The random fading $h$ is often modeled as a (circularly symmetric complex)
Gaussian random variable,
though this is not fundamental for the exposition here.
More importantly, however, it is usually assumed that the fading $h$
is known exactly to the receiver, which is motivated by
signal measurements that can be acquired at the receiver.
We will also make this assumption throughout this paper.
A natural follow-up question concerns whether $h$ is also known to the
transmitter. In this paper, we will generally assume that the transmitter is
{\em ignorant} of $h.$

\subsubsection{Multiple signals interfere in a linear additive way}
This second point is a direct extension of the first one.
Namely, consider now that $L$ transmitters are active simultaneously.
Then, along the same lines described above, the 
induced signal at any point in space can be expressed as $\sum_{\ell = 1}^{L} {h_{\ell} X_\ell[t]}.$
It is commonly assumed that the respective fading coefficients $h_{\ell}$
are independent of each other, each following a Gaussian law,
and we will follow this assumption throughout, although it is not
fundamental for our main arguments.

\subsubsection{Noise is independent of the signal and added at the receiver}
More particularly, in line with the standard models, we will assume that the
noise distribution is described by a (circularly symmetric complex) Gaussian law, although again this is not fundamental
for the ideas laid out here.

These three key observations, together with several more detailed considerations,
lead to the following commonly used model for the signal at a particular receiver
when $L$ transmitters are simultaneously active:
\begin{align}
Y[t] = \sum_{\ell = 1}^{L} {h_{\ell} X_\ell[t]} + Z[t] \ .  \label{e:wireless}
\end{align}
As this model shows, the undesirable element for linear network coding on the physical layer is the noise: Generally, it will add up over the various stages of the network, and thus, suitable reliable coding is necessary.

\subsection{Gaussian Channel Capacity} \label{s:gaussian}

Consider the special case of the model above where there is only one active transmitter. This is the classical Gaussian channel:
\begin{align}
Y[t] = hX[t] + Z[t]
\end{align} if $Z[t]$ is taken to be independent and identically distributed (i.i.d.) circularly symmetric complex Gaussian noise with variance $\sigma^2$. To model the fact that the transmitter has a limited power budget, it is usually required that the transmitted signal satisfies
\begin{align}
\frac{1}{n} \sum_{t=1}^n{|X[t]|^2} \leq P \ .
\end{align} The seminal paper of Shannon showed that the capacity of this channel is
\begin{align}
C = \log_2\left(1 + \frac{|h|^2 P}{\sigma^2}\right)
\end{align} bits per channel use \cite{shannon48}. Of course, Shannon only showed the existence of good block codes, not any explicit constructions. After more than sixty years of research, codes with low complexity encoding and decoding algorithms have been developed that can come quite close to the capacity. Describing these codes is far beyond the scope of this paper so we point the interested reader to a survey by Forney and Costello that appeared in an earlier issue of these Proceedings \cite{fc07}. For our purposes, we will assume the existence of good encoders and decoders that can achieve the channel capacity. One important aspect of a capacity-achieving code is that the elements of each codeword look as if they were sampled i.i.d. from a Gaussian distribution with variance $P$. In Section \ref{s:codes}, we will describe some very recent efforts to design practically viable codes for physical layer network coding. 

\section{Uncoded Strategies} \label{s:uncoded}

A quick comparison of (\ref{e:netcod}) and (\ref{e:wireless}) reveals that our wireless channel model has an input-output relationship that is nearly the same as our desired network coding operation.
Specifically, both the wireless channel and random linear network coding output a linear combination of their inputs, with the coefficients generated randomly according to some distribution. However, the wireless channel exhibits two key differences:
\begin{enumerate}
\item It operates over the \textit{complex field} instead of a finite field.
\item The receiver only observes a \textit{noisy} version of the linear combination. 
\end{enumerate} Our goal, for the remainder of this paper, is to show that by using appropriate modulation and coding techniques we can exploit the wireless medium for reliable network coding over a finite field. In this section, we will see how much is already possible using \textit{uncoded} modulation strategies. In other words, we will try to map the complex field into the finite field but we will ignore the effects of the noise.

\subsection{Finite Constellations} \label{s:uncodedfinite}

The most intuitive physical layer network coding strategy is to have the users transmit their message bits directly on the wireless channel. If the channel gains are equal, then the receiver will get the noisy sum of the bits from which it can make an estimate of its desired modulo-$2$ sum. To the best of our understanding, this key idea was independently and concurrently proposed by three research groups in 2006: Zhang, Liew, and Lam \cite{zll06}, Popovski and Yomo \cite{py06ICC}, and ourselves \cite{ng06}. In their paper, Zhang, Liew, and Lam also coined the term physical layer network coding. Here, we examine this strategy in the context of the two-way relay channel.

Consider the two-way relay channel in Figure \ref{f:twowaychanmodel} and assume that both users transmit simultaneously. Ideally, we would like the channel to directly compute the mod-$2$ sum of the transmitted bits. The relay could then broadcast the sum back to the users and complete the entire exchange in only two time slots. Of course, the channel does not output the mod-$2$ sum so we will have to be a bit more clever. For now, assume that each user knows its channel gain to the relay and can invert it. The relay therefore sees the noisy sum of the transmitted signals and we can consider the real and imaginary parts of the channel separately. The real part of the signal observed at the relay at time $t$ is
\begin{align}
\Yr[t] = X_1[t] + X_2[t] + \Zr[t]
\end{align} where $X_\ell[t]$ is the real-valued symbol transmitted by user $\ell$ at time $t$ and $\Zr[t]$ is Gaussian noise with variance $\sigma^2$. We now examine a simple strategy for transmitting a noisy modulo-$2$ sum of the bits to the relay over the real part of the channel. The same scheme can be applied to the imaginary part.

\begin{figure}[h]
\centering
\includegraphics[width=3.25in]{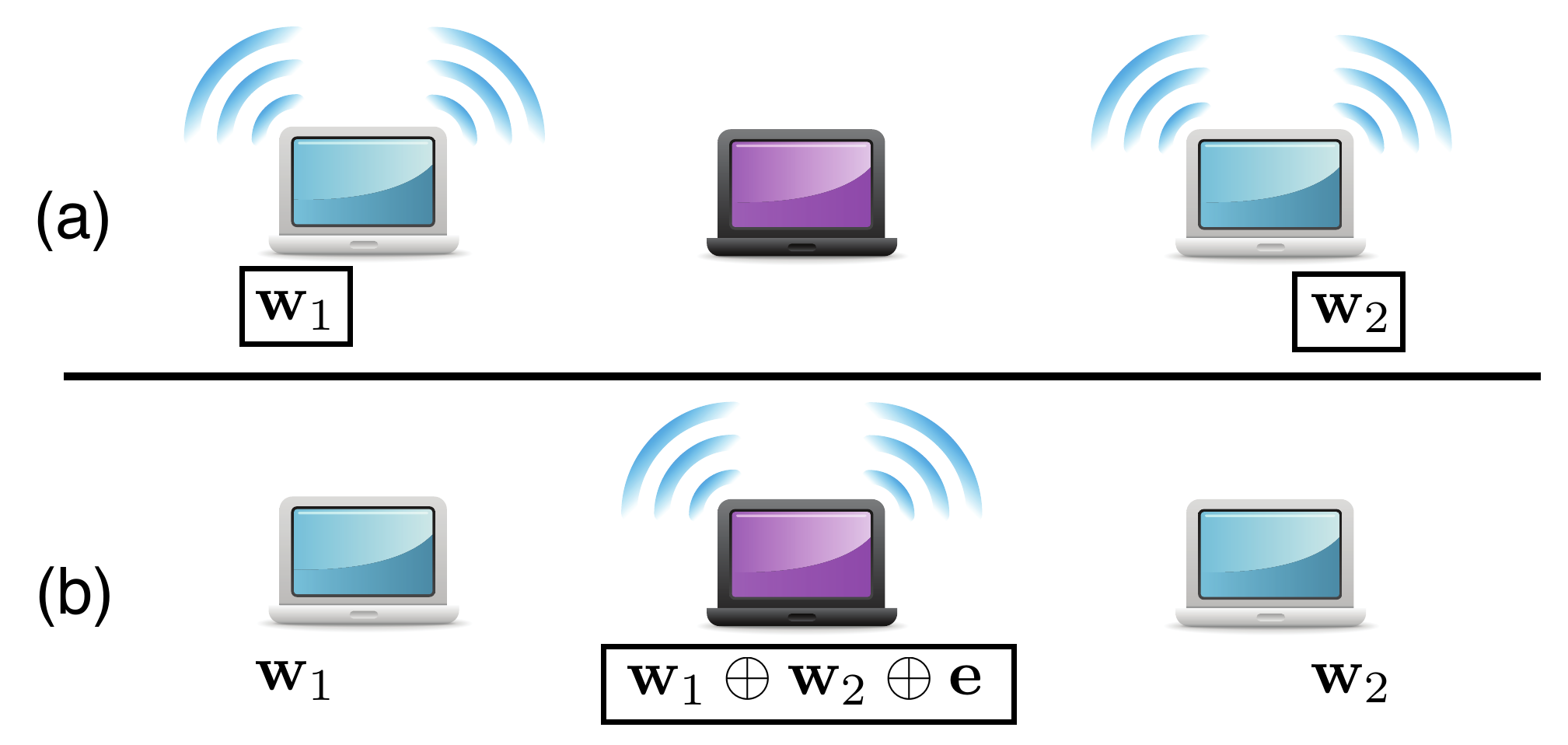}
\caption{A physical layer network coding strategy for the two-way relay channel that requires $2$ time slots. (a) During the first time slot, both users transmit their messages which gives the relay access to a noisy sum of the packets, $\mathbf{w}_1 \oplus \mathbf{w}_2 \oplus \mathbf{e}$. (b) During the second time slot, the relay broadcasts this corrupted sum of the messages back to both users. They use knowledge of their own message to infer a corrupted version of the other user's message. With an end-to-end error correcting code, this scheme can be used for reliably exchanging information.}\label{f:twowayphysical}
\end{figure}

For ease of analysis, assume that the total power per channel use is $P=2$ which means that we can allocate one unit of power to the real part and one unit to the imaginary part.\footnote{Note that we can model any signal-to-noise ratio by changing the noise variance.} Let $W_\ell$ denote the bit from user $\ell$. Each user maps its bit to a channel input symbol using binary phase-shift keying (BPSK)
\begin{align}
X_\ell = \begin{cases}
1 & W_\ell = 1,\\
-1 & W_\ell = 0.
\end{cases}
\end{align} Therefore, if the mod-$2$ sum of the bits $U = W_1 \oplus W_2$ is $1$, then the sum of the transmitted signals $X_1 + X_2 = 0$. Similarly, if $U$ is $0$, then $X_1 + X_2$ is either $2$ or $-2$, depending on the original bits. We would like to design a decoding rule for the delay to make an estimate $\hat{U}$ of the mod-$2$ sum $U$ from its noisy observed sum $Y$. For simplicity, we assume that each user's bit is generated from a fair coin toss. The maximum a posteriori (MAP) rule to minimize the probability that $\hat{U}$ is in error is given by:
\begin{align}
\hat{U}_{\text{MAP}} = \argmax_{b=0,1} f_Y(y | U = b) \Pr(U = b)
\end{align} where $f_Y(y | U = b)$ is the conditional probability distribution of the channel output given the mod-$2$ sum. Since the noise is Gaussian, the probability density function is
\begin{align}
&f_{\Yr}(y | U = b) = \\ &~\begin{cases} \frac{1}{\sqrt{2\pi \sigma^2}} e^{-y^2/2\sigma^2} & U = 1, \\
\frac{1}{2\sqrt{2\pi \sigma^2}} \left(e^{-(y-2)^2/2\sigma^2} +e^{-(y+2)^2/2\sigma^2}\right) & U = 0.
\end{cases} \nonumber 
\end{align} Note that for $U = 0$, $\Yr$ follows a Gaussian mixture distribution since $X_1 + X_2$ can be either $2$ or $-2$ with equal probability. A bit of calculation reveals that the MAP rule is just a threshold on the magnitude of the received signal
\begin{align}
\hat{U}_{\text{MAP}} = \begin{cases}
1 & |\Yr| \leq 1 + (\sigma^2 \ln 2)/2,\\
0 & \mbox{otherwise.} 
\end{cases}
\end{align} and the error probabilities can be computed using the Q-function. Applying this strategy to every bit allows the relay to obtain a corrupted version of the modulo-$2$ sum of the bit strings,
\begin{align}
\mathbf{\hat{u}} = \mathbf{w}_1 \oplus \mathbf{w}_2 \oplus \mathbf{e}
\end{align} where $\mathbf{e}$ is the error vector. In Figure \ref{f:twowayphysical}, we illustrate how this can enable more efficient communication over the two-way relay channel. In one time slot, both transmitters send their messages concurrently, giving the relay the noisy modulo-$2$ sum. In the next time slot, the relay broadcasts the modulo-$2$ sum to the users, which can then solve for a corrupted version of their desired bits. Coupling this strategy with an end-to-end error correcting code allows the users to successfully exchange messages using just $2$ time slots. While this coarse analysis seems to favor uncoded transmission, we must account for the level of error correction needed to recover from the errors introduced at the relay. In Section \ref{s:performance}, we compare the performance of this strategy to other strategies. 

There has been a great deal of interest in the idea behind this example and many powerful extensions and generalizations have been developed for fading channels \cite{hgdtl07,kpt08,chk09,kimphd, ast10}, asynchronous scenarios \cite{zll06ITW,rz09,wfl09}, code division multiple-access \cite{cy09}, and practical deployments \cite{kgk07}. See \cite{pk09} for a survey. As mentioned earlier, the signal observed at the relay can be treated as the output of a noisy modulo-$2$ adder so the relay can attempt to denoise the modulo-$2$ sum if each transmitter employs the same linear code \cite{ng06, py06ICC, zzll07}. In practice, this can be accomplished using low-complexity codes such as fountain codes \cite{pljw08}, repeat-accumulate codes\cite{zl09B}, or low-density parity-check codes \cite{ltxw10}. In larger networks, errors can also be left uncorrected and dealt with using the end-to-end network error correction framework proposed in \cite{kk08}.

\subsection{Analog Signaling}\label{s:analogsignal}

Instead of mapping the complex-valued output of the wireless channel into a finite field, the linear network coding framework can be modified to operate directly in the complex field. This can result in significant performance gains as the desired linear combinations are identical to the operation performed by the channel, ignoring the noise. Thus, if the signal-to-noise-ratio ($\snr$) is sufficiently high, this ``analog'' strategy should perform quite well. This approach, often called amplify-and-forward \cite{sg00,gv02,ltw04,ehm07,bzg07}, was proposed for two-way relaying in 2006 by Popovski and Yomo \cite{py06VTC} as well as Rankov and Wittneben \cite{rw06}. Here, we examine how this strategy can be applied to the two-way relay channel.

The channel from the users to the relay is given by
\begin{align}
\Yr[t] = X_1[t] + X_2[t] + \Zr[t]
\end{align} where $X_\ell[t]$ is the complex symbol transmitted by user $\ell$ at time $t$ and $\Zr[t]$ is circularly symmetric complex Gaussian noise with variance $\sigma^2$. Recall that each user must meet its power constraint, $\frac{1}{n}\sum_{t=1}^n{|X_\ell[t]|^2} \leq P$. Each user encodes its message $\mathbf{w}_\ell$ into a codeword $\mathbf{x}_\ell$ using a capacity-achieving code for a single user Gaussian channel as in Section \ref{s:gaussian}. The symbols of $\mathbf{x}_\ell$ are i.i.d. according to a Gaussian distribution with variance $P$. If we assume the messages are independent, then the codewords are also independent and the observed vector at the relay $\yr$ is the sum of independent Gaussian vectors. The variance of $\Yr[t]$ is $2P + \sigma^2$. As desired, the relay now has a noisy sum of the transmitted signals which it can broadcast back to the users. The channels from the relay to users $1$ and $2$ are
\begin{align}
Y_1[t] &= \Xr[t] + Z_1[t] \\
 Y_2[t] &= \Xr[t] + Z_2[t]
\end{align} with $\Xr[t]$ as the symbol transmitted by relay at time $t$ and $Z_\ell[t] \sim \mathcal{CN}(0,\sigma^2)$. We assume that equal amounts of time are devoted to sending and receiving. The relay simply retransmits its noisy sum $\Yr[t]$, scaled to meet the power constraint,
\begin{align}
\Xr[t] &= \sqrt{\frac{P}{2P + \sigma^2}} \Yr[t].
\end{align} Each user then observes an even noisier version of the sum,
\begin{align}
Y_1[t] &= \sqrt{\frac{P}{2P + \sigma^2}} \bigg(X_1[t] + X_2[t] + \Zr[t]\bigg)+ Z_1[t] \\
Y_2[t] &= \sqrt{\frac{P}{2P + \sigma^2}} \bigg(X_1[t] + X_2[t] + \Zr[t]\bigg) + Z_2[t],
\end{align} from which it can subtract its own signal and obtain a corrupted version of the signal transmitted by the other user. The $\snr$ of the resulting channel is
\begin{align}
\frac{P}{\sigma^2} \left(\frac{P}{3P + \sigma^2}\right)
\end{align} which means that each user can (theoretically) sustain a rate up to 
\begin{align}
R_{\text{ANALOG}} = \frac{1}{2} \log\left(1 + \frac{P}{\sigma^2} \left(\frac{P}{3P + \sigma^2}\right)\right) \label{e:analog}
\end{align} bits per channel use while keeping the probability of error arbitrarily small. Note that the factor of $\frac{1}{2}$ comes from using one time slot to communicate to the relay and another to communicate back to the users. At high $\snr$, this is quite close to the ideal performance which is the rate achievable by one user communicating via the relay as if the other was silent. This rate is 
\begin{align}
R_{\text{UPPER}} = \frac{1}{2} \log\left(1 + \frac{P}{\sigma^2}\right) \label{e:upper}
\end{align} bits per channel use and can serve as an upper bound on our schemes.

For comparison, a routing strategy requires $4$ time slots (as discussed in Section \ref{s:twoway}) and can only achieve a rate of 
\begin{align}
R_{\text{ROUTING}} = \frac{1}{4} \log\left(1 + \frac{P}{\sigma^2}\right) \label{e:routing}
\end{align} bits per channel use. If the relay performs network coding on its received packets, then $3$ time slots are required and an achievable rate of
\begin{align}
R_{\text{NETCOD}} = \frac{1}{3} \log\left(1 + \frac{P}{\sigma^2}\right) \label{e:netcod}
\end{align} bits per channel use is possible.\footnote{If we vary the ratio of time spent sending to receiving, then it is theoretically possible to reach slightly higher rates.}

In Section \ref{s:performance}, the rate curves of these schemes as well as the BPSK scheme in Section \ref{s:uncodedfinite} and the lattice scheme in Section \ref{s:equalgains} are plotted and compared.

In general, the wireless channel may be subject to fading and the users and relay may have more than one antenna. This scenario has been studied in detail in the literature \cite{hkezwb07,vh08,uk08,zlcc09,js10,amz10}. For larger networks, the approach remains the same: each relay scales and retransmits its observation \cite{kmgkm07}. However, the noise build ups with each retransmission so the $\snr$ requirement increases as the network grows \cite{bzg07}. We also note that, in some scenarios, it is advantageous to use the compress-and-forward framework from \cite{lkec10} instead of amplify-and-forward.

\subsection{Cross-Layer Design}

While analog network coding presents clear benefits in terms of end-to-end throughput, these seem to come at an \textit{architectural} price. Today's wireless networks use a layered architecture \cite{clcd07} (often referred to as the network protocol stack) that separates wireless signaling schemes from flow control, scheduling, and information contents.
In order to make this decoupling possible, the physical layer employs error-correcting codes
to lower the raw error probabilities, and on top of that, an acknowledgment feedback process (often expressed
as ACK/NACK) is used according to which a transmitter repeats a certain packet until it
receives an acknowledgment signal from the receiver.
Higher layers then rely on an error-free transmission of a certain capacity. Indeed, this capacity
limitation is the only way in which higher layers are aware of the underlying physical reality.
With analog network coding, it is not possible to enable this.
Here, intermediate nodes will forward erroneous packets (or functions thereof), and there is
no way for an intermediate node to tell whether it is forwarding something useful or not.
For reliable communication, additional error correction has to be implemented end-to-end.
This may be acceptable for small networks (such as the two-way relay channel example),
but it will prevent a layered architecture for larger networks,
instead requiring cross-layer design.
By this, we mean that higher layers have to take into account a detailed description of the physical reality of the communications medium, beyond simple capacity figures. Kawadia and Kumar argued that the benefits of cross-layer design may be offset by the loss of robustness and modularity in \cite{kk05}.

In Sections \ref{s:reliablepnc} and \ref{s:fading}, we suggest a framework for reliable physical layer network coding across wireless links. This framework can be implemented in a completely modular fashion since each receiver can reliably decode a linear function of the transmitted bits. Standard acknowledgment feedback protocols can be employed in this scenario, as well as novel feedback protocols developed with network coding in mind \cite{ssm09}. Network layer protocols can be built around state-of-the-art wired network coding algorithms. This framework has shown promise in several theoretical studies but much work remains to show that it can be successfully adapted to practice.

\section{Reliable Physical Layer Network Coding} \label{s:reliablepnc}

The framework developed in Section \ref{s:finitefield} is able to protect linear combinations of codewords against noise. However, the tools seem to only be a natural fit for channels that can be written as a noisy operation over a finite field. While this behavior can be mimicked on a wireless channel by using appropriate hard decision rules (as in Section \ref{s:uncodedfinite}), this approach does not, in general, lead to the best performance. In this section, we describe a principled approach to wireless channels with equal channel gains. Ultimately, this approach will enable us to build a digital interface for physical layer network coding over channels with arbitrary gains. 

\subsection{Nested Lattice Codes}

A natural starting point is to find a capacity-achieving code for the Gaussian channel with a \textit{linear structure}. Specifically, we would like a good code over the reals such that the sum of any two codewords is itself a codeword. It turns out that this matches the definition of a \textit{lattice}. A lattice $\Lambda$ is a set of real-valued vectors (or points in $\mathbb{R}^n$) such that for any two elements $\lambda_1, \lambda_2 \in \Lambda$ we have that $\lambda_1 + \lambda_2 \in \Lambda$.\footnote{We will only employ lattices that contain the zero vector $\mathbf{0}$ so that if $\lambda \in \Lambda$ then $-\lambda \in \Lambda$ as well.} For example, one simple lattice is just the set of all integers, $\mathbb{Z}$. Of course, a lattice, by itself, cannot be used as a codebook as it contains an infinite number of points and clearly violates the power constraint. A lattice code is usually constructed by carving out a portion of a lattice and designating the selected points as codewords. Several such constructions have been proposed that can achieve the capacity of a Gaussian channel (see, for instance, \cite{debuda89,lsz93,poltyrev94,loeliger97,ur98,ftc00,zse02,ez04,elz05,ecd04}).

The lattice code should also obey some form of modulo arithmetic so that we can map between the linear combination taken by the channel and our desired linear combination over the messages. One elegant solution is to employ the capacity-achieving \textit{nested lattice codes} developed by Erez and Zamir \cite{ez04}. If a lattice $\Lambda$ is a subset of another lattice $\Lambda_{\text{FINE}}$, $\Lambda \subset \Lambda_{\text{FINE}}$, then $\Lambda$ is said to be nested in $\Lambda_{\text{FINE}}$. In this nested lattice pair, $\Lambda$ is often referred to as the coarse lattice and $\Lambda_{\text{FINE}}$ as the fine lattice. One simple example is the integer lattice $\mathbb{Z}$ taken together with the integer multiples of $q$, $q\mathbb{Z}$. Let $Q(\cdot)$ be a quantizer that maps vectors to the nearest lattice point in Euclidean distance:
\begin{align}
Q_{\Lambda}(\mathbf{x}) = \argmin_{\lambda \in \Lambda} \| \mathbf{x} - \lambda \| \ .
\end{align}
The set of points that quantize to a given lattice point are called the Voronoi region of that point.  The Voronoi region for the zero vector is referred to as the fundamental Voronoi region,  
\begin{align}
\mathcal{V}_{\Lambda} = \{ \mathbf{x} : Q_{\Lambda}(\mathbf{x}) = \mathbf{0}\} \ .
\end{align} Note that, for a lattice, each Voronoi region is just a translation of the fundamental Voronoi region.

A nested lattice code $\mathcal{L}$ is the set of points of the fine lattice that lie within the fundamental Voronoi region $\mathcal{V}_{\Lambda}$ of the coarse lattice, $\mathcal{L} = \Lambda_{\text{FINE}} \cap \mathcal{V}_\Lambda$. For our example, with the integers and integers multiplied by $q$, the resulting nested lattice code is just $\{0,1,2,\ldots, q-1\}$. Erez and Zamir's nested lattice codes have three key properties that make them ideally suited for reliable physical layer network coding:
\begin{enumerate}
\item As the dimension (or blocklength) increases, the fundamental Voronoi region of the coarse lattice becomes more spherical, which means that it is appropriate for guaranteeing the power constraint.
\item As the dimension increases, the Voronoi regions of the fine lattice also become more spherical, which means that they afford good protection against Gaussian noise.
\item The underlying lattices are built from codes over a finite field, which means that finite field messages can be mapped onto the codebook and back without sacrificing linearity.
\end{enumerate} For more details, see the excellent overview by Zamir, Shamai, and Erez \cite{zse02} as well as \cite{ez04,elz05,zamir09}.

The modulo operation for a nested lattice code is defined to be the quantization error,
\begin{align}
[\mathbf{x}] \modl = \mathbf{x} - Q_{\Lambda}(\mathbf{x}),
\end{align} and can be shown to satisfy the distributive property,
\begin{align}
\Big[ [\mathbf{x}_1] \modl + \mathbf{x}_2 \Big] \modl = [\mathbf{x}_1 + \mathbf{x}_2] \modl, 
\end{align} and the commutative property with respect to quantization onto the fine lattice,
\begin{align}
\left[Q_{\Lambda_{\text{FINE}}}\left([\mathbf{x}] \modl\right)\right] \modl = \left[Q_{\Lambda_{\text{FINE}}}\left(\mathbf{x}\right)\right] \modl .
\end{align}

The encoding and decoding algorithms for the Erez-Zamir scheme are easy to describe. First, the encoder maps its message $\mathbf{w}$ onto a point in the nested lattice code $\mathbf{x} \in \mathcal{L}$ and sends it across the channel. The decoder observes the transmitted codeword in noise $\mathbf{y} = \mathbf{x} + \mathbf{z}$ and makes the following estimate:
\begin{align}
\mathbf{\hat{x}} = \left[Q_{\Lambda_{\text{FINE}}}(\alpha \mathbf{y})\right] \modl. \label{e:latticedecoder}
\end{align} In words, the decoder first scales its observed vector by $\alpha$, quantizes onto the fine lattice, and then takes the modulus (to ensure that the decoded fine lattice point is really in the nested code). It can be shown that (for long enough blocklengths $n$) the estimate $\mathbf{\hat{x}}$ is equal to the transmitted codeword $\mathbf{x}$ with high probability so long as the rate is at most \begin{align}
R = \frac{1}{2} \log_2\left(\frac{P}{N_{\text{EFFEC}}}\right) \label{e:nestedrate}
\end{align} where we define the effective noise variance to be
\begin{align}
N_{\text{EFFEC}} = \frac{1}{n} \| \alpha \mathbf{y} - \mathbf{x} \|^2. 
\end{align} Taking $\alpha = 1$, we get that the effective noise variance is just $\sigma^2$, the variance of the channel noise. Unfortunately, this does not take us all the way to the channel capacity, only to $\frac{1}{2}\log_2\left(\frac{P}{\sigma^2}\right)$. Erez and Zamir showed that to reach capacity, it is crucial that the channel observation be scaled by the minimum-mean squared error (MMSE) coefficient, $\alpha = \frac{P}{P + \sigma^2}$ (see \cite{forney03} for an in-depth discussion of the MMSE scaling). Now, the effective noise variance is
\begin{align}
N_{\text{EFFEC}} &= \frac{1}{n} \| \alpha (\mathbf{x} + \mathbf{z}) - \mathbf{x} \|^2 \\
&= \frac{1}{n} \| (\alpha - 1) \mathbf{x} + \alpha \mathbf{z} \|^2 \\
&= (\alpha - 1)^2 P + \alpha^2 \sigma^2 \\
&= \frac{P\sigma^2}{P + \sigma^2} 
\end{align} where the second to last step is, roughly speaking, due to the fact that capacity-achieving codes have codewords that look like i.i.d. Gaussian vectors.\footnote{The observant reader will have noticed that part of the noise comes from the codeword itself. Proving that this ``self-noise'' is only as bad as additional Gaussian noise is technically subtle and usually involves the use of dithering vectors that are removed prior to decoding. This analysis is beyond the scope of this survey and we refer curious readers \cite{ez04} for more details.} Plugging this into the rate expression in (\ref{e:nestedrate}) yields $\frac{1}{2}\log_2\left(1 + \frac{P}{\sigma^2}\right)$. Note that here we have only dealt with a real-valued Gaussian channel. Repeating the same scheme over the imaginary part of the channel yields the full capacity of a complex-valued Gaussian channel.

This scheme can be adapted to the two-way relay channel in a straightforward fashion. Before doing so, we mention some guidelines for employing nested lattice codes in practice. These guidelines are drawn from \cite{zse02,ez04,et05} and we refer interested readers to these works for more detail. First, the coarse lattice can simply be taken to be integers multiplied by the field size $q$, $\Lambda = q\mathbb{Z}^n$. This reduces quantization onto the coarse lattice to rounding each element of the received vector to the nearest multiple of $q$. The cost is the shaping gain, which is at most $0.509$ bits per complex channel symbol (see \cite{ez04}). Next, the fine lattice $\Lambda_{\text{FINE}}$ can be the codewords of any linear code over the finite field of size $q$. Any popular low-complexity linear code can be employed, such as a low-density parity check (LDPC) code. Finally, the fine lattice quantizer can be replaced with an appropriate low-complexity decoding algorithm (written as $\textsc{Decoder}$ below). The cost of replacing the fine lattice with a low-complexity code is just the gap to capacity for this code. Mathematically, the encoder just becomes multiplication by the generator matrix, $\mathbf{x} = \mathbf{Gw}$. The decoding operation in (\ref{e:latticedecoder}) simplifies to
\begin{align}
\mathbf{\hat{x}} &= \left[\textsc{Decoder}( \alpha \mathbf{y})\right] \modqz.
\end{align} Thus, through the use of modern low-complexity codes, nested lattice codes can be implemented in a practically feasible manner. See \cite{et05} for more details. In Section \ref{s:codes}, we give an overview of some very recent efforts to build low-complexity codes that are specifically tailored for reliable physical layer network coding.

\subsection{Equal Channel Gains}\label{s:equalgains}

We now show that the relay in the Gaussian two-way relay channel can recover the sum of the transmitted codewords modulo the coarse lattice, $\mathbf{x}_{\text{SUM}} = [ \mathbf{x}_1 + \mathbf{x}_2 ]\hspace{0.05in}\modl$. This nested lattice scheme was proposed by Narayanan, Wilson, and Sprintson for the two-way relay channel in 2007 \cite{nws07, wnps10}. Concurrently, we proposed a lattice framework for general Gaussian multiple-access networks that attains the same performance for the two-way relay channel \cite{ng07allerton,ng08ETT}. Subsequently, Nam, Chung, and Lee proposed a nested lattice scheme for unequal (but known) channel gains \cite{ncl08,ncl10}.

Each user encodes its message onto a codeword from a good nested lattice code. Recall that the relay observes the sum of the channel inputs plus Gaussian noise, $\yr = \xb_1 + \xb_2 + \zr$. The relay uses the decoding rule 
\begin{align}
\mathbf{\hat{x}}_{\text{SUM}} &= \left[ Q_{\Lambda_{\text{FINE}}}(\alpha \yr)\right] \modl \\ & = \left[ Q_{\Lambda_{\text{FINE}}}([\alpha \yr] \modl )\right] \modl 
\end{align} where the last line is due to the commutative property of the modulo operation. We now show that the term inside the quantization operator is just the desired sum $\mathbf{x}_{\text{SUM}}$ plus some noise,
\begin{align}
&[\alpha \yr] \modl \\
&= [ \alpha( \mathbf{x}_1 + \mathbf{x}_2 + \zr)] \modl \\
&= [ \mathbf{x}_1 + \mathbf{x}_2 + (\alpha - 1) (\mathbf{x}_1 + \mathbf{x}_2) + \alpha \zr] \modl \\ & = [ [\mathbf{x}_1 + \mathbf{x}_2] \modl + (\alpha - 1) (\mathbf{x}_1 + \mathbf{x}_2) + \alpha \zr] \modl \nonumber \\ & = [ \mathbf{x}_{\text{SUM}}+ (\alpha - 1) (\mathbf{x}_1 + \mathbf{x}_2) + \alpha \zr] \modl, 
\end{align} where the second-to-last step follows by the distributive property of the modulo operation. Therefore, the effective noise variance is 
\begin{align}
N_{\text{EFFEC}} &= \frac{1}{n} \| (\alpha - 1) (\mathbf{x}_1 + \mathbf{x}_2) + \alpha \zr  \|^2 \\ &= (\alpha - 1)^2 2P + \alpha^2 \sigma^2
\end{align} where the last line is due to the fact that $\mathbf{x}_1$ and $\mathbf{x}_2$ are independent vectors which (nearly) look as if drawn from an i.i.d. Gaussian distribution with power $P$. The optimal scaling coefficient is the MMSE coefficient $\alpha = \frac{2P}{2 P + \sigma^2}$ and plugging this into (\ref{e:nestedrate}) yields a rate of $\frac{1}{2}\log{\left(\frac{1}{2} + \frac{P}{\sigma^2}\right)}$. If this scheme is used on both the real and imaginary dimensions, the relay can decode the sum at any rate up to
\begin{align}
R_{\text{COMP}} =  \log_2\left(\frac{1}{2} + \frac{P}{\sigma^2}\right).
\end{align} 

\begin{figure*}[!ht]
\centering
\includegraphics[width=7in]{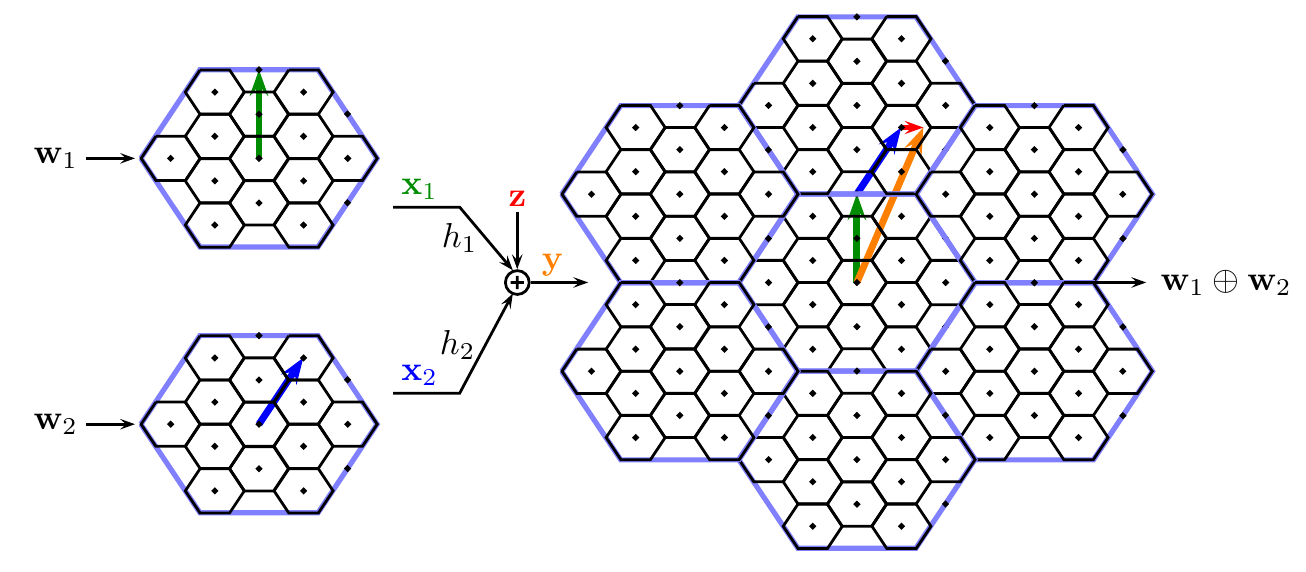}
\caption{Each transmitter maps its finite field message into an element of the nested lattice code and sends this vector on the channel. Here, the channel coefficients are taken to be equal, $h_1,h_2=1$. Therefore, the receiver observes a noisy sum of the transmitted vectors and determines the closest lattice point. After taking a modulo operation with respect to the coarse lattice, the receiver can invert the mapping and determine the modulo sum of the original messages.} \label{f:latticecompcode}
\end{figure*}

In Figure \ref{f:latticecompcode}, the nested lattice scheme described above is illustrated for the special case of $\alpha = 1$. Each user transmits a vector from the nested lattice code. The channel outputs the noisy sum, which is observed at the receiver. Note that the sum of the two vectors exceeds the boundary of the original nested lattice code. However, by decoding to the closest fine lattice point and then taking the modulo operation, the modulo sum of the codewords can be recovered.

Recall that the ultimate goal is for the relay to reliably decode the modulo sum of the original messages, not the codewords. While this is not necessarily possible for \textit{any} nested lattice code, the Erez-Zamir codes used here are constructed using a linear code over a finite field. It can therefore be shown that there exists a mapping $\phi$ from the finite field message vectors onto the nested lattice code that preserves linearity. See Lemma 6 in our recent paper for a construction of $\phi$ \cite{ng09IT}. Mathematically, this property can be expressed as 
\begin{align}
\mbox{Encoding:}~~&\mathbf{x}_\ell = \phi( \mathbf{w}_\ell) \\
\mbox{Decoding:}~~&\phi^{-1}\Big( \left[a_1 \mathbf{x}_1 + a_2 \mathbf{x}_2 + \cdots + a_L \mathbf{x}_L\right] \modl \Big) \nonumber \\&=  a_1 \mathbf{w}_1 \oplus a_2 \mathbf{w}_2 \oplus \cdots \oplus a_L \mathbf{w}_L \ .
\label{e:nestedlinear}
\end{align} For the low-complexity case where the coarse lattice is $q \mathbb{Z}^n$ and the fine lattice is a linear code, $\phi$ is just the generator matrix $\mathbf{G}$ of the linear code and $\phi^{-1}$ is its inverse.

This mapping is the last piece of the puzzle. With it, the sum of the messages can be recovered directly from the modulo sum of the codewords,
\begin{align}
\phi^{-1}\Big( [\mathbf{x}_1 + \mathbf{x}_2] \modl \Big) = \mathbf{w}_1 \oplus \mathbf{w}_2.
\end{align} Now, we can use this in a two-way communication scheme by using one time slot to transmit the sum of the messages to the relay and another to send it back to the users. It follows that the users can exchange messages at any rate up to 
\begin{align}
R_{\text{LATTICE}} = \frac{1}{2} \log_2\left(\frac{1}{2} + \frac{P}{\sigma^2}\right). \label{e:twowaylattice}
\end{align} This rate nearly matches the upper bound in (\ref{e:upper}) except for a missing $\frac{1}{2}$ inside the logarithm.\footnote{Several groups have unsuccessfully tried to find a lattice scheme that can attain the upper bound. This remains an open problem.}  

This two-way lattice scheme has been extensively studied and generalized in the literature. These extensions include unequal channel gains\cite{ncl10,bc08}, non-Gaussian channel models \cite{ez08}, secret messages \cite{hy09}, private messages \cite{hgs09}, direct links \cite{sd10}, as well as more than two transmitters \cite{gsgps09,sakh10,okj10}. Gupta-Kumar style scaling laws \cite{gk00} have also been derived for this lattice scheme \cite{lfqc09}. We also note that similar lattice-based schemes can increase achievable rates in interference channels \cite{bpt10,sjvjs08}.  

Overall, this nested lattice scheme can be used as a digital framework for physical layer network coding on the wireless channel. It is able to exploit the addition performed by the channel while preserving modulo arithmetic and protecting against Gaussian noise.  In a larger network, each relay will recover a linear combination of the original messages. It can then transmit this linear combination as its own message, just as relays in wireline networks send out linear combinations of their received messages. In Section \ref{s:fading}, we will generalize the results in this section to unequal channel gains. Furthermore, we show that the transmitters do not even need to know the channel gains, which means that this scheme can be applied to fading channels and scenarios with more than one receiver. In the next section, we plot the performance of each scheme discussed so far for the Gaussian two-way relay channel.

\section{Performance Comparison}\label{s:performance}

\begin{figure}[h]
\centering
\includegraphics[width=3.5in]{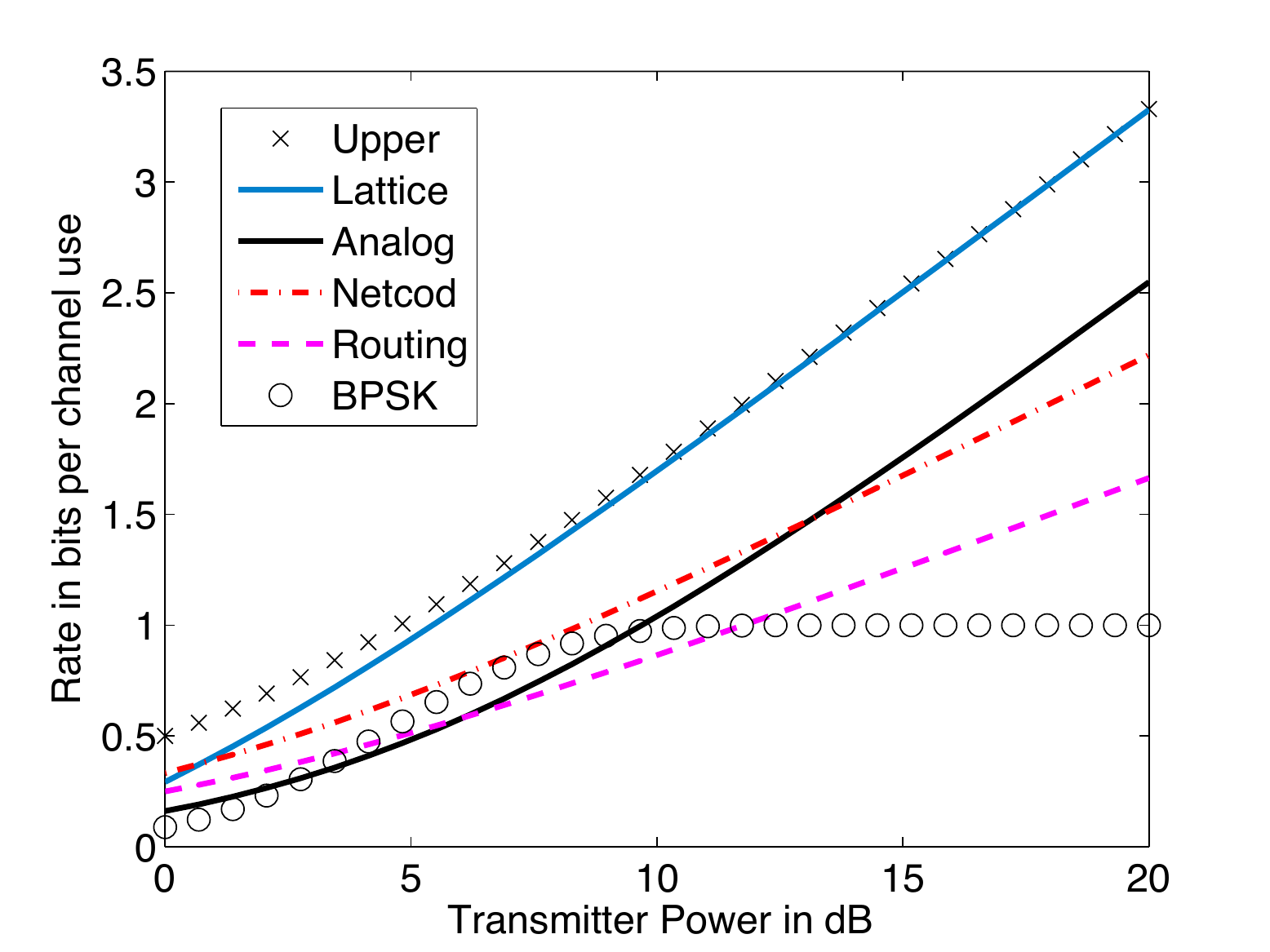}
\caption{A performance comparison of the schemes for the two-way relay channel discussed in this paper. }\label{f:twowayplot}
\end{figure}

In Figure~\ref{f:twowayplot}, we compare the performance for the various network coding strategies discussed
in the present paper, for the particular case of a Gaussian two-way relay channel. The figure displays the rate per user in bits per channel use, as a function of the transmit power per user, while the noise is assumed to be of unit variance. Starting from the top, the figure shows the simple upper bound given in Equation (\ref{e:upper}). It is instructive to consider the behavior at large transmit power $P$, characterized by the limit of the ratio $R / \log(1 + P/\sigma^2).$ For the upper bound, it is clear that this limiting slope is $1/2.$

The next curve, labeled ``Lattice,'' is the performance of reliable physical layer network coding via lattice codes,
given in Equation (\ref{e:twowaylattice}). We note that this scheme is close to the upper bound and that it attains the same limiting slope of $1/2.$

The following curve, labeled ``Analog,'' represents the analog network coding scheme discussed in Section~\ref{s:analogsignal}. It follows the upper bound but never meets it. This is because the noise observed at the relay is sent along with the desired signal, an effect that would be even more detrimental if there were further stages in the network. In the limit of high transmit power $P,$ this effect becomes negligible and the optimal limiting
slope of $1/2$ is attained.

The curve labeled ``Netcod'' is the performance attained by the wireless broadcast network coding scheme in Section \ref{s:twoway}, depicted in Figure \ref{f:twowaynetcod}. Each user takes a turn sending its message to the relay and the relay sends the mod-$2$ sum back to the users. This scheme loses out at high transmit power due to the fact that three channel uses are needed for each exchange, making for a limiting slope of $1/3.$

The curve labeled ``Routing'' is the scheme in Section \ref{s:twoway}, depicted in Figure \ref{f:twowayrouting}. Each user takes a turn sending its message to the relay and the relay sends these back to the users. At high transmit power, one can verify that this leads to a limiting slope of $1/4.$

The final curve, labeled ``BPSK,'' is the binary scheme in Section \ref{s:uncodedfinite}, according to which each user transmits its bits uncoded and the relay makes a hard decision about the modulo-$2$ sum. The broadcast phase is abstracted as a bit pipe simultaneously to both users with a rate corresponding to the capacity of the broadcast channel from the relay to the users (which thus depends on the transmit power $P$).
Error-correcting codes are then used end-to-end. Note that due to the fact that BPSK is used, this scheme plateaus at $1$ bit per channel use; if a larger constellation were used, this plateau effect would be higher (but the performance at low SNR might suffer).

As a final caveat, we note that some of the achievable schemes can be slightly improved by optimizing the ratio of time spent in the different phases. For the present figure, it is assumed that all these phases are of the same length, as in the descriptions provided in Section \ref{s:twoway}.

\section{Fading Channels} \label{s:fading}

If the channel simply outputs a noisy sum of the transmitted signals, then it is intuitive that this can be exploited for adding up the messages. Yet, in general, the channel output will be some linear combination according to complex-valued coefficients and it is not immediately clear that this will be a good match for network coding over a finite field. Here, we demonstrate how to overcome this obstacle using the compute-and-forward framework we proposed in \cite{ng09IT}. First, we show how to reliably compute over real-valued channels and then we use this scheme as a building block for complex-valued channels.

\subsection{Real-Valued Channels}

Consider a real-valued channel whose output vector is just a linear function of the transmitted vectors plus some Gaussian noise,
\begin{align}
\mathbf{y} = h_1 \mathbf{x}_1 + h_2 \mathbf{x}_2 + \cdots + h_L \mathbf{x}_L + \mathbf{z}.
\end{align} Assume that each user selects and transmits a point from a nested lattice code. The key idea is that, instead of trying to decode the sum, the receiver should aim to decode an \textit{integer combination} of the codewords (modulo the coarse lattice), 
\begin{align}
\mathbf{v} = [ a_1 \mathbf{x}_1 + a_2 \mathbf{x}_2 + \cdots + a_L \mathbf{x}_L ] \modl.
\end{align} This integer combination is itself a codeword, due to the linear structure of the nested lattice code, and is therefore afforded protection against noise. If these integer coefficients are close enough to the real-valued coefficients of the channel, then it seems plausible that the receiver can decode the function successfully. More precisely, the receiver makes the following estimate of $\mathbf{v}$:
\begin{align}
\mathbf{\hat{v}} &= \left[ Q_{\Lambda_{\text{FINE}}}(\alpha \yb)\right] \modl \\ & = \left[ Q_{\Lambda_{\text{FINE}}}([\alpha \yb] \modl )\right] \modl 
\end{align} where the second line is due to the commutative property. Prior to quantization onto the fine lattice, this is just the desired function $\mathbf{v}$ plus some noise:
\begin{align}
&[\alpha \yb] \modl \\ &= [ \alpha( h_1 \mathbf{x}_1 + \cdots + h_L \mathbf{x}_L + \zb)] \modl \\
&= [ a_1\mathbf{x}_1 +\cdots + a_L\mathbf{x}_L ~\cdots \\ & ~~\cdots~ + (\alpha h_1 - a_1) \mathbf{x}_1 +\cdots  +(\alpha h_L - a_L) \mathbf{x}_L + \alpha \zb] \modl \nonumber \\ & = [\mathbf{v} + (\alpha h_1 - a_1) \mathbf{x}_1 +\cdots  (\alpha h_L - a_L) \mathbf{x}_L + \alpha \zb] \modl \nonumber.
\end{align} The effective noise comes from both the non-integer part of the channel and the Gaussian noise. The effective noise variance is 
\begin{align}
N_{\text{EFFEC}} &= \frac{1}{n} \|\alpha \mathbf{z}+ (\alpha h_1 -a_1)\mathbf{x}_1 + \cdots + (\alpha h_L - a_L) \mathbf{x}_L  \|^2 \nonumber \\
&= \alpha^2 \sigma^2 + P \sum_{\ell = 1}^L (\alpha h_\ell - a_\ell)^2.
\end{align} This means that the receiver can recover the integer combination of codewords so long as the rate of the nested lattice code is at most
\begin{align}
R_{\text{COMP}} = \frac{1}{2} \log_2 \left( \frac{P}{\alpha^2 \sigma^2 + P \sum_{\ell = 1}^L (\alpha h_\ell - a_\ell)^2}\right).
\end{align} Again, it can be shown that the optimal $\alpha$ is the MMSE coefficient, 
\begin{align}
\alpha_{\text{MMSE}} = \frac{P\sum_{\ell=1}^L{h_\ell a_\ell}}{\sigma^2 + P \sum_{\ell= 1}^L |h_\ell|^2}.
\end{align} Here, the role of $\alpha$ can be thought of as trying to move the channel coefficients towards integers. For instance, if the channel coefficients are $h_1 = 0.5$ and $h_2 = 0.5$, then choosing $\alpha = 2$ converts the channel into a noisy adder (like that studied in Section \ref{s:equalgains}) at the price of quadrupling the noise variance. Substituting $\alpha_{\text{MMSE}}$ into the rate expression simplifies it  down to
\begin{align}
R_{\text{COMP}} = \frac{1}{2} \log_2\left( \left(\sum_{\ell=1}^L{| a_\ell |^2} - \alpha_{\text{MMSE}} \sum_{\ell = 1}^L{h_\ell a_\ell} \right)^{-1} \right) \label{e:compopt}
\end{align} Finally, the recovered integer combination of the codewords can be mapped to a modulo-$q$ linear combination of the messages, 
\begin{align}
\phi^{-1}( \mathbf{v}) = a_1 \mathbf{w}_1 \oplus a_2 \mathbf{w}_2 \oplus \cdots \oplus a_L \mathbf{w}_L.  
\end{align}

\begin{figure}[h]
\centering
\includegraphics[width=3.5in]{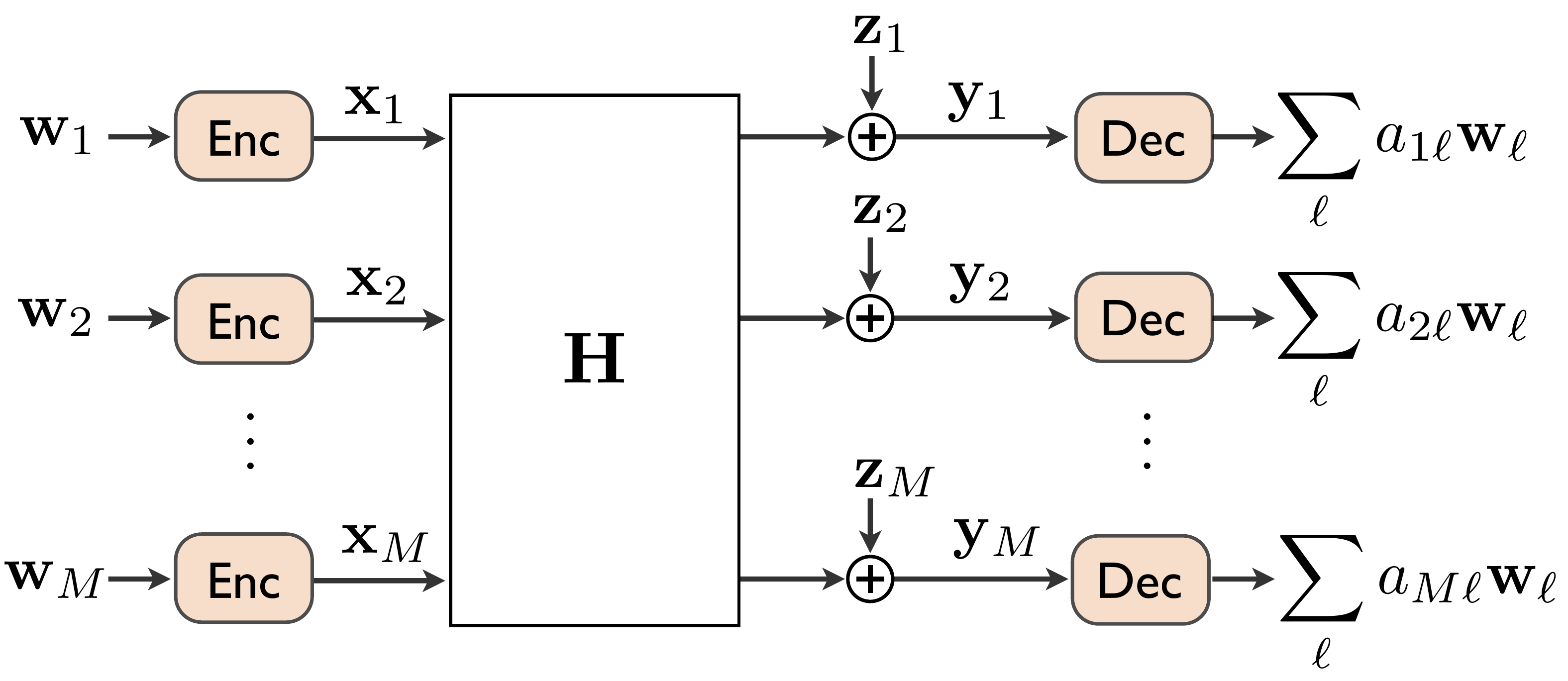}
\caption{In a wireless network, each receiver is free to decode the linear combination of messages that best fits its observed channel coefficients.}\label{f:multirec}
\end{figure}

Therefore, this channel model can be exploited to reliably compute linear functions over a finite field. We now draw attention to the fact that the encoding function does not depend on the channel coefficients: the transmitters just send out their nested lattice point. This implies that even if the channel coefficients are completely unknown to the transmitters, this scheme can still be used. (Clearly, the receiver needs to know the channel coefficients so it can choose the integer coefficients appropriately and determine the optimal $\alpha$.) Thus, if multiple receivers with different channel coefficients all receive signals from the same transmitters, each receiver can decode its own particular linear combination at a rate as given in (\ref{e:compopt}). See Figure \ref{f:multirec} for an illustration.

Note that the receiver is not limited to decoding a single function. It can run the decoding step on its observed vector several times to extract several different functions. As a special case, the receiver can recover a single message $\mathbf{w}_m$ by setting the associated coefficient to one, $a_m = 1$, and all others to zero. It can be shown that the rate for this special case is
\begin{align}
R = \frac{1}{2} \log_2{\left(1 + \frac{Ph_m^2}{\sigma^2 + P\sum_{\ell \neq m} h_\ell^2}\right)} \label{e:interferenceasnoise}
\end{align} which corresponds exactly to the rate attainable by treating other simultaneous transmissions as undesirable interference. Taking this idea a bit further, it can be demonstrated that the nested lattice framework can attain any point in the Gaussian multiple-access rate region \cite{ng09IT}. Thus, it can only help to use this nested lattice scheme for decoding functions since, as a backup, the receiver can just decode the messages individually with no rate penalty whatsoever.

Unlike the finite field strategies considered in Section \ref{s:finitefield}, different coefficients result in different rates. In many scenarios,  the receiver can simply search for the coefficients that offer the highest possible rate and then target that equation. If the channel coefficients are sufficiently independent, then there is a good chance that each receiver will decode a linearly independent equation and the end-to-end linear transformation between the sources and the destinations will be invertible. In some cases, there are benefits to imposing some restrictions on the coefficients that can be selected. 

\begin{figure}[h]
\centering
\includegraphics[width=2.75in]{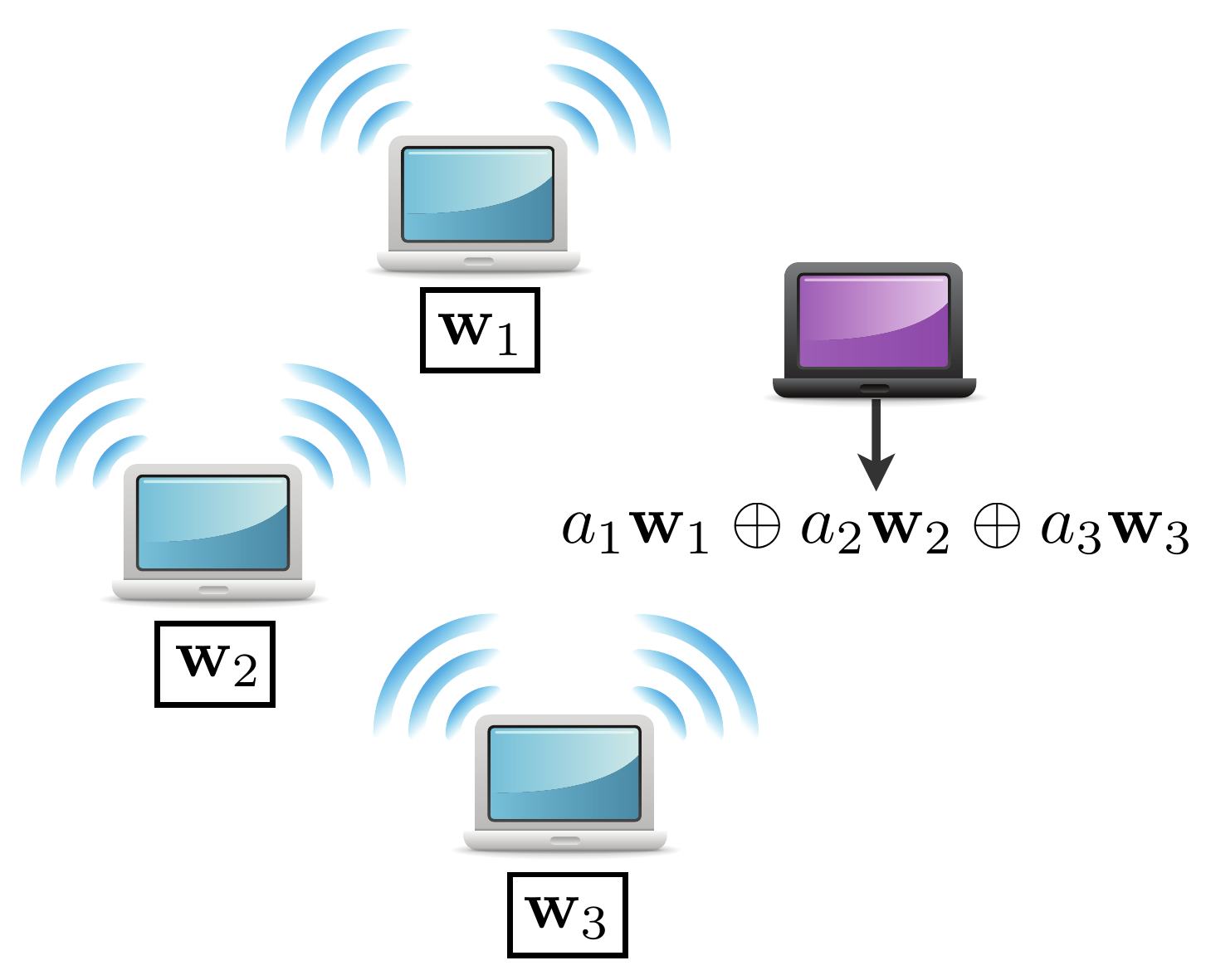}
\caption{Three users simultaneously transmit their messages to a relay that wishes to decode a linear function, $a_1 \mathbf{w}_1 \oplus a_2 \mathbf{w}_2 \oplus a_3 \mathbf{w}_3$. In Figure \ref{f:geteqm3}, three strategies for this scenario are plotted.}\label{f:m3compute}
\end{figure}

\begin{figure}[h]
\centering
\includegraphics[width=3.5in]{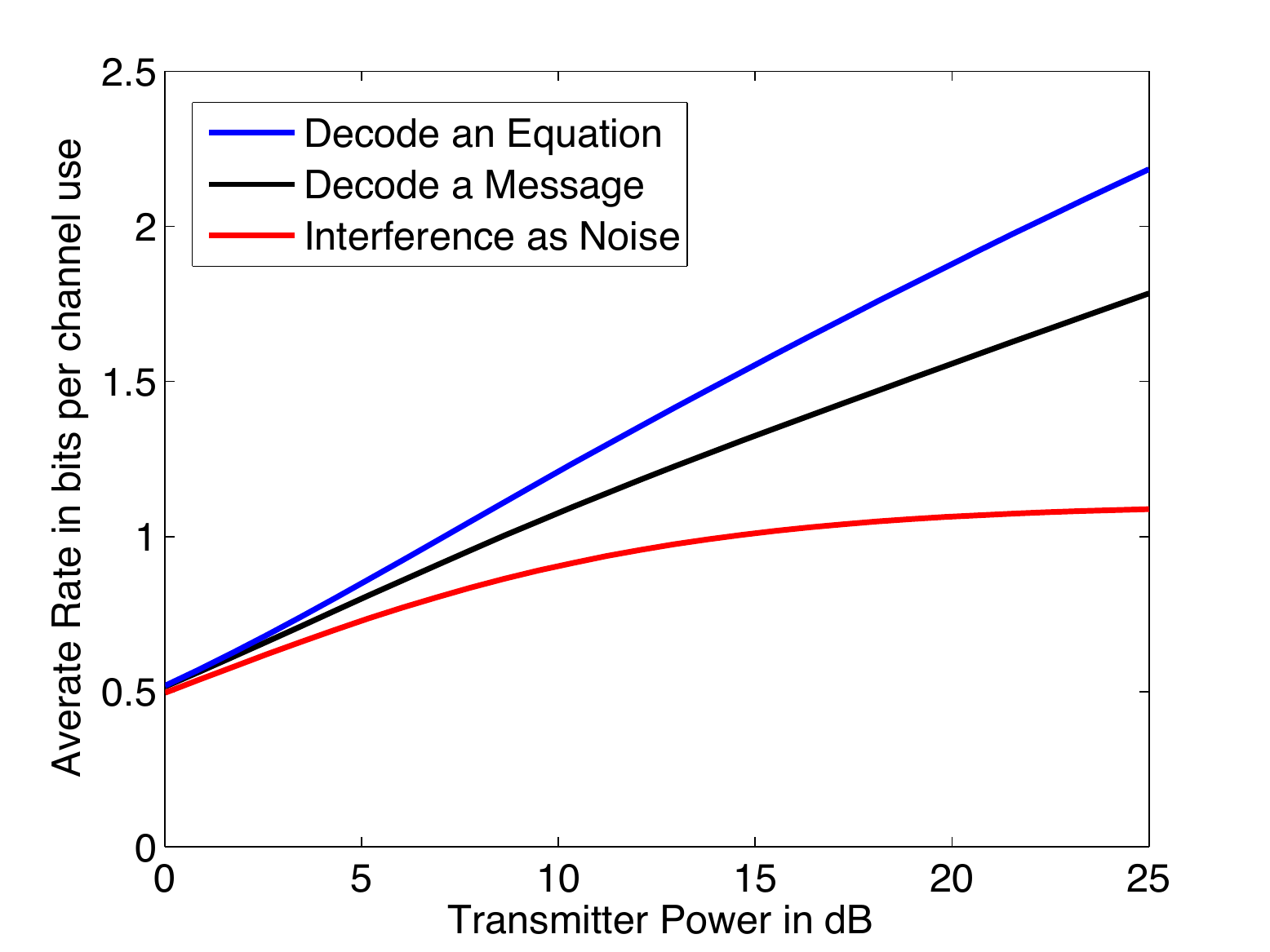}
\caption{Average rate for recovering an equation of three simultaneously transmitted messages. The top line corresponds to using a nested lattice code to decode an equation. The middle line corresponds to the information-theoretic optimum for recovering a single message. The bottom line corresponds to recovering a single message while treating the other two as noise. The fading is drawn from a Gaussian distribution and is only known to the receiver.}\label{f:geteqm3}
\end{figure}

Overall, this strategy is especially useful in networks with concurrent transmissions as receivers must cope with interference in one way or another. As an example, in Figure \ref{f:m3compute}, there are three transmitters that simultaneously transmit their messages to a single receiver which tries to reliably decode a linear function. Note that recovering a single message corresponds to recovering a function with a single non-zero coefficient. Each fading coefficient is drawn i.i.d. from a Gaussian distribution and the fading realization is only known to the transmitter. In Figure \ref{f:geteqm3}, we compare three different strategies for this network. The ``Decode an Equation'' strategy is just the nested lattice scheme discussed in this section, whose performance is given by (\ref{e:compopt}). The ``Decode a Message'' strategy is the rate attainable if the receiver only tries to get a single message, not an equation. This curve corresponds to the information-theoretic optimum for the special case of decoding a single message. Finally, the ``Interference as Noise'' is a suboptimal strategy for decoding one message that treats the other two messages as noise. The performance is given by (\ref{e:interferenceasnoise}). 

\subsection{Complex-Valued Channels} 

As mentioned in Section \ref{s:wireless}, the baseband representation, 
\begin{align}
\mathbf{y} = h_1 \mathbf{x}_1 + h_2 \mathbf{x}_2 + \cdots + h_L \mathbf{x}_L + \mathbf{z},
\end{align}of a narrowband wireless channel is complex-valued. With a few simple modifications, the nested lattice scheme for real-valued channels can be applied here as well. Each encoder breaks its message $\mathbf{w}_\ell$ into two messages of equal length, $\mathbf{w}_\ell^{\Re}$ and $\mathbf{w}_\ell^{\Im}$. The resulting messages are mapped onto nested lattice points (using $\phi$) and sent as the real and imaginary parts of the transmitted codeword,
\begin{align}
\Re(\mathbf{x}_\ell) = \phi(\mathbf{w}_\ell^{\Re}) \qquad \qquad \qquad \Im(\mathbf{x}_\ell) = \phi(\mathbf{w}_\ell^{\Im}). 
\end{align} The decoder deals with the real and imaginary parts of the received signal separately. The real part is
\begin{align}
\Re(\mathbf{y}) = \sum_{\ell=1}^L\Re(h_\ell) \Re(\xb_\ell) - \Im(h_\ell) \Im(\xb_\ell) + \Re(\zb)
\end{align} which can be treated like the received signal from a real-valued channel with $2L$ transmitters. From this, the receiver can decode a linear function of the form
\begin{align}
\mathbf{u}^{\Re} = \left[ \sum_{\ell=1}^L a_\ell^{\Re} \wb_\ell^{\Re} -a_\ell^{\Im}  \wb_\ell^{\Im}\right]\modq.
\end{align} Similarly, the imaginary part is 
\begin{align}
\Im(\mathbf{y}) =  \sum_{\ell=1}^L\Im(h_\ell) \Re(\xb_\ell) + \Re(h_\ell) \Im(\xb_\ell) + \Im(\zb)
\end{align} from which the receiver recovers the complementary linear function
\begin{align}
\mathbf{u}^{\Im} = \left[ \sum_{\ell=1}^L a_\ell^{\Im} \wb_\ell^{\Re} + a_\ell^{\Re}  \wb_\ell^{\Im}\right]\modq.
\end{align} In \cite{ng09IT}, we showed that the effective noise variance is always the same for real and imaginary received signals and the computation rate
\begin{align}
R_{\text{COMP}} = \log_2\left(\frac{P}{| \alpha |^2 \sigma^2 + P \sum_{\ell=1}^L{ |\alpha h_\ell - a^{\Re}_\ell - j a^{\Im}_\ell|^2}}\right) \nonumber
\end{align} bits per channel use is achievable. If a destination is given several of these equations, it can infer the original messages as long as the matrix of complex integer coefficients
\begin{align}
\mathbf{A} = \left[
\begin{array}{cccc}
  a_{11}^{\Re} + j a_{11}^{\Im}& a_{12}^{\Re} + j a_{12}^{\Im}  & \cdots &a_{1L}^{\Re} + j a_{1L}^{\Im} \\
   a_{21}^{\Re} + j a_{21}^{\Im}& a_{22}^{\Re} + j a_{22}^{\Im}  & \cdots &a_{2L}^{\Re} + j a_{2L}^{\Im}  \\
  \vdots & \vdots  & \ddots & \vdots \\
  a_{M1}^{\Re} + j a_{M1}^{\Im}& a_{M2}^{\Re} + j a_{M2}^{\Im}  & \cdots &a_{ML}^{\Re} + j a_{ML}^{\Im}
\end{array} 
\right] \nonumber
\end{align} is full rank. This scheme is studied in detail in \cite{ng09IT} as well as in the first author's PhD thesis \cite{nazerphd}. One interesting extension is to use multiple antennas at the receiver to steer the channel coefficients towards integer values, thus increasing the rate at which an equation can be recovered \cite{znge09}.

\section{Code Constructions} \label{s:codes}

The above analysis shows that, for sufficiently large blocklengths, there exist encoders and decoders that can efficiently harness the interference property of the wireless medium for network coding. While we have chosen codebooks with a linear structure, this does not, by itself, imply they can be implemented with low complexity, especially on the decoder side. This is similar to the classical channel coding problem where the capacity gives the ultimate performance limit and one then attempts to design practically realizable codes that can approach this limit. Very recently, several groups have proposed practical coding schemes that are designed with reliable physical layer network coding in mind. We briefly describe three of these schemes below.

In \cite{fsk10}, Feng, Silva, and Kschischang take an algebraic approach and propose a set of lattice partitions whose properties make them appealing for physical layer network coding. For instance, their nested lattice construction attains a field size of $q^2$ for the same complexity it takes the basic scheme to attain a field size of $q$. Through simulations, they also show that their framework works quite well for blocklengths as small as $100$.

The basic compute-and-forward framework employs constellations and codes over a prime-sized finite field to ensure a match between the linear combinations taken by the channel and the linear combinations over the codewords. Unfortunately, the implementation complexity of a coding scheme increases with the characteristic\footnote{The size of any finite field can be written as $q^K$ where $q$ is a prime and $K$ is a positive integer. Usually, $q$ is referred to as the characteristic of the finite field.} of the finite field. Therefore, in practice, it is desirable to have constellations of size $2^K$ for some positive integer $K$. With an appropriate mapping, these constellations can be coupled with binary linear codes to significantly reduce the implementation complexity. The main difficulty is that the mapping from constellation points to codeword symbols must preserve the interference structure of the channel. To overcome this issue, Hern and Narayanan proposed using multilevel codes and allowing the receiver to decode to a larger class of functions \cite{hn10}. Specifically, each constellation symbol is specified by a block of $K$ bits. Different coefficients are allowed for each of these bits but are constant from symbol to symbol. This allows the receiver more freedom in mapping the channel operation to something useful over the finite field. Independently, Ordentlich and Erez have developed a framework \cite{oe10} that uses mapping by set partitioning to go from binary codewords to higher order constellations. Their preliminary simulations have shown that this constellation mapping in conjuction with an LDPC code can perform quite well.

\section{Larger Networks}
 
In the context of a larger network, what does reliable physical layer network coding mean for the overall network code? Network codes are usually designed over networks of bit pipes and the relays are therefore free to select any coefficients for their linear combinations that maximize the end-to-end throughput. However, in this new framework, the rate at which a particular linear combination can be decoded by a relay hinges on the match between the desired coefficients and the fading coefficients of the wireless channel. Relays must choose function coefficients that balance between maximizing the local computation rate and the end-to-end throughput. For instance, in the two-way relay channel with fading, the relay can recover any equation $a_1 \mathbf{w}_1 \oplus a_2 \mathbf{w}_2$ so long as both $a_1$ and $a_2$ are non-zero. Otherwise, at least one of the users will not receive any novel information. 

For a multi-stage relay network, selecting the optimal function coefficients corresponds to an integer program if we are given access to the full channel state information. Yet, in many cases, nodes will only have access to some subset of the channel state, most likely just that of their immediate neighbors \cite{aas10}. For this scenario, new algorithms and heuristics are needed to enable distributed coefficient selection. More research is also needed to elucidate how much of an advantage, in terms of end-to-end throughput, is offered by reliable physical layer network coding over realistic wireless network topologies. One preliminary theoretical analysis is developed in~\cite{ggw10}, where it is assumed that any node receives the modulo-2 sum of the packets transmitted by the nearest-neighbor transmitters. This situation is compared to the case where nodes can receive full packets from neighboring nodes, but at an appropriately lower rate. In terms of transport capacity and for a network of nodes located on a regular lattice in two dimensions, it is shown that the former more than doubles the capacity of the latter.

\section{Conclusions}
Physical layer network coding is an intuitively pleasing enhancement to network coding. In network coding, intermediate nodes forward linear combinations of information packets (or more general functions of them) and destination nodes collect a sufficient number of linearly independent functions so as to be able to recover their desired information packets. Physical layer network coding starts from the insight that it can be much more efficient for a node to directly learn the linear function of several packets, rather than having to first learn each packet separately and then evaluating the function. This is particularly tempting in the case of the wireless medium since there, transmitted signals naturally interfere in a linear fashion. One approach reviewed in this paper, dubbed analog network coding, uses this effect in an uncoded fashion, dealing separately (end-to-end) with the accumulating noise. In this paper, we particularly emphasized a way to circumvent this problem and instead enforce {\em reliable} physical layer network coding even inside of the network. The main idea underlying this framework is that each transmitter should employ the same linear code. Through examples, we have illustrated the superiority of this approach in a capacity sense. A second superiority concerns the resulting overall layered system architecture: In reliable physical layer network coding, the physical layer can be made essentially transparent to the end-to-end communication process.

Systems are currently being designed to determine to what extent the promised performance gains are attainable over the real wireless medium, but much of this exciting path still lays ahead.

\section*{Acknowledgments}

The authors would like to thank G. Bresler, U. Erez, K. Narayanan, G. Reeves, A. D. Sarwate, S. Shamai, M. Wilson, R. Zamir, and J. Zhan for valuable discussions and N. Devroye for a careful reading of the manuscript. They would also like to thank the reviewers for suggestions that helped improve the presentation of this paper.

\bibliographystyle{IEEEtran}
%\bibliography{procbib}

% Generated by IEEEtran.bst, version: 1.13 (2008/09/30)

\begin{IEEEbiography}[{\includegraphics[width=1in,height=1.25in,clip,keepaspectratio]{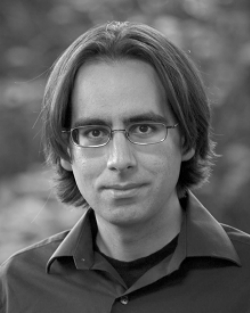}}]{Bobak Nazer} received the B.S.E.E. degree from Rice University, Houston, TX, in 2003, the M.S. degree from the University of California, Berkeley, CA, in 2005, and the Ph.D degree from the University of California, Berkeley, CA, in 2009, all in electrical engineering. 

He is currently an Assistant Professor in the Department of Electrical and Computer Engineering at Boston University, Boston, MA. From 2009 to 2010, he was a postdoctoral associate in the Department of Electrical and Computer Engineering at the University of Wisconsin, Madison, WI. His research interests are in information theory, communications, and signal processing, with a focus on developing new techniques for distributed, reliable computation over networks.
 
Prof. Nazer won the Eli Jury Award for his dissertation research from the Department of Electrical Engineering and Computer Sciences at the University of California, Berkeley, CA. He is a member of Eta Kappa Nu, Tau Beta Pi, and Phi Beta Kappa. 
\end{IEEEbiography}

\begin{IEEEbiography}[{\includegraphics[width=1in,height=1.25in,clip,keepaspectratio]{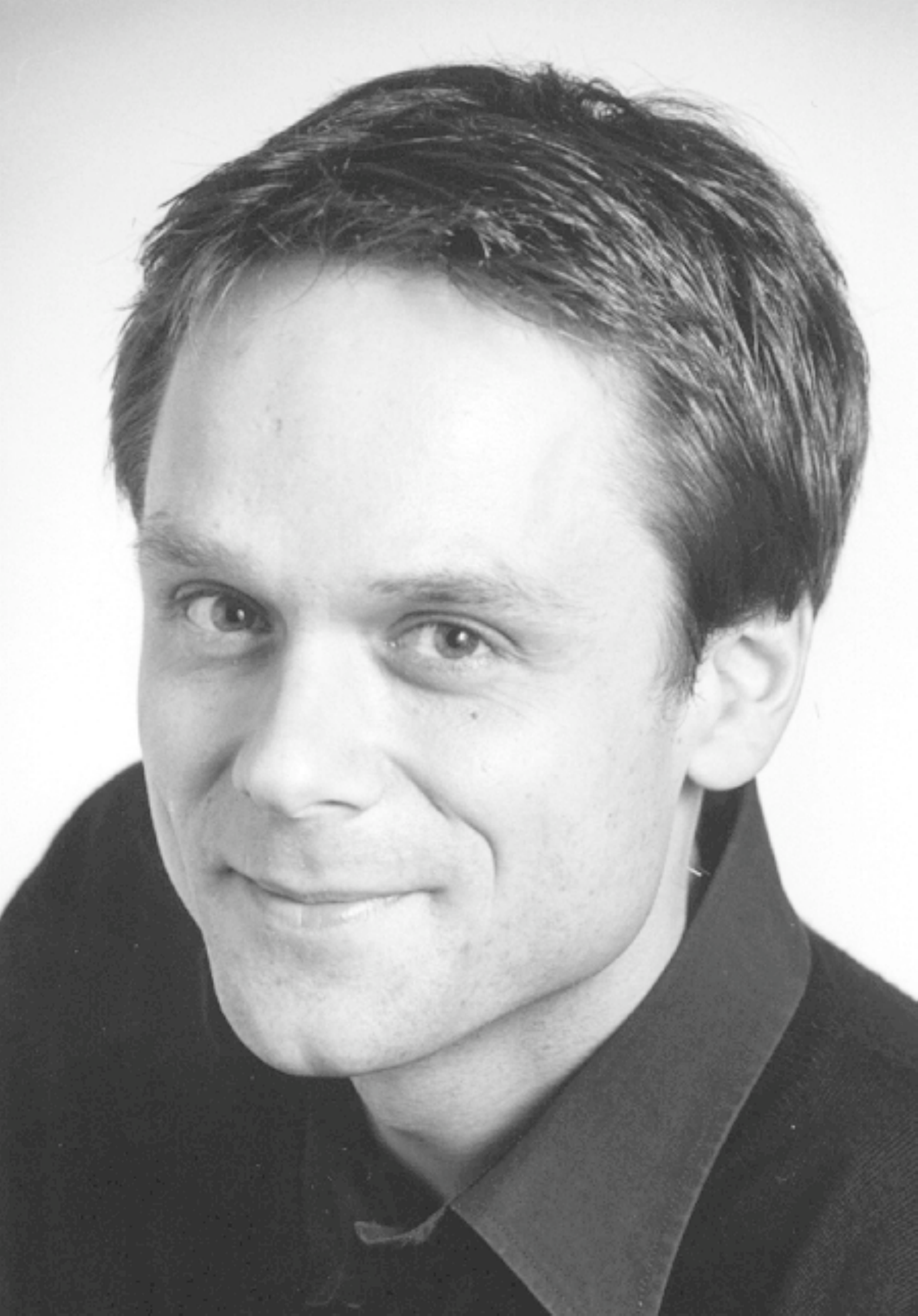}}]{Michael Gastpar} received the Dipl. El.-Ing. degree from the Swiss Federal Institute of Technology (ETH), Zurich, in 1997, the M.S. degree from the University of Illinois at Urbana-Champaign, Urbana, in 1999, and the
Doctorat \`es Science degree from the Swiss Federal Institute of Technology (EPFL), Lausanne, in 2002, all in electrical engineering. He was also a student in engineering and philosophy at the Universities of Edinburgh and Lausanne.

He is currently an Associate Professor with the Department of Electrical Engineering and Computer Sciences, University of California, Berkeley, and a Professor with the Department of Electrical Engineering, Mathematics, and Computer Science, Delft University of Technology, Delft, The Netherlands. He was a researcher with the Mathematics of Communications Department, Bell Labs, Lucent Technologies, Murray Hill, NJ. His research interests are in network information theory and related coding and signal processing techniques, with applications to sensor networks and neuroscience.

Prof. Gastpar won the 2002 EPFL Best Thesis Award, an NSF CAREER award in 2004, and an Okawa Foundation Research Grant in 2008. He is an Information Theory Society Distinguished Lecturer (2009--2010). He is currently an Associate Editor for Shannon Theory for the IEEE TRANSACTIONS ON INFORMATION THEORY, and he has served as Technical Program Committee Co-Chair for the 2010 International Symposium on Information Theory, Austin, TX. \end{IEEEbiography}

\end{document}